\documentclass{aa}

\usepackage{graphicx,xcolor,natbib}
\usepackage{txfonts}
\usepackage{soul}
\usepackage{array}
\usepackage{lscape}

\begin{document}

   \title{Chemical composition of planetary hosts}  
   \subtitle{C, N, and $\alpha$-element abundances 
   \thanks{Based on stellar observations obtained with the 1.65 m telescope and VUES spectrograph at the Mol\.{e}tai Astronomical Observatory, Institute of Theoretical Physics and Astronomy, Vilnius University.}
   \thanks{Full Tables~\ref{table:Results} and \ref{table:exoplanets} are only available in electronic form at the CDS via anonymous ftp to cdsarc.u-strasbg.fr () or via \url{https://cdsarc.cds.unistra.fr/viz-bin/qcat?J/A+A/}}
   }

   \author{A. Sharma
           \inst{1},          
           E. Stonkut\.{e}
           \inst{1},
           A. Drazdauskas
           \inst{1},
           R. Minkevi\v{c}i\={u}t\.{e}
           \inst{1},
           \v{S}. Mikolaitis
           \inst{1},
           G. Tautvai\v{s}ien\.{e}
           \inst{1},
           \and
           T. Narbuntas
           \inst{1}
          }

   \institute{Institute of Theoretical Physics and Astronomy, Vilnius University, Sauletekio av. 3, 10257, Vilnius, Lithuania\\
              \email{ashutosh.sharma@ff.stud.vu.lt}
             }

   \date{}

   \titlerunning{Chemical composition of planetary hosts. C, N, and $\alpha$-element abundances}
   \authorrunning{Sharma et al.}

 
  \abstract
   {Accurate atmospheric parameters and chemical composition of planet hosts play a major role in characterising exoplanets and understanding their formation and evolution.}
   {Our objective is to uniformly determine atmospheric parameters and chemical abundances of carbon (C), nitrogen (N), oxygen(O), and the $\alpha$-elements, magnesium (Mg) and silicon (Si), along with C/O, N/O and Mg/Si abundance ratios for planet-hosting stars. In this analysis, we aim to investigate the potential links between stellar chemistry and the presence of planets.}
   {Our study is based on high-resolution spectra of 149 F, G, and K dwarf and giant stars hosting planets or planetary systems. The spectra were obtained with the Vilnius University Echelle Spectrograph (VUES) on the 1.65 m Mol\.{e}tai Observatory telescope. The determination of stellar parameters was based on a standard analysis using equivalent widths and one-dimensional, plane-parallel model atmospheres calculated under the assumption of local thermodynamical equilibrium. The differential synthetic spectrum method was used to uniformly determine the carbon C(C2), nitrogen N(CN), oxygen [O I], magnesium Mg~I, and silicon Si~I elemental abundances as well as the C/O, N/O, and Mg/Si ratios.} 
   {We analysed elemental abundances and ratios in dwarf and giant stars, finding that [C/Fe], [O/Fe], and [Mg/Fe] are lower in metal-rich dwarf hosts; whereas [N/Fe] is close to the Solar ratio. Giants show smaller scatter in [C/Fe] and [O/Fe] and lower than the Solar average [C/Fe] and C/O ratios. The (C+N+O) abundances increase with [Fe/H] in giant stars, with a minimal scatter. We also noted an overabundance of Mg and Si in planet-hosting stars, particularly at lower metallicities, and a lower Mg/Si ratio in stars with planets. In giants hosting high-mass planets, nitrogen shows a moderate positive relationship with planet mass. C/O and N/O ratios show moderate negative and positive slopes in giant stars, respectively. The Mg/Si ratio shows a negative correlation with planet mass across the entire stellar sample.}
   {}


   \maketitle

\section{Introduction} \label{sec:intro}

The study of planets outside the Solar system recently has developed rapidly in the field of astrophysics. Thanks to ground and space missions such as Kepler \citep{Kepler10}, Transiting Exoplanet Survey Satellite (TESS) \citep{Rinehart15}, over 5\,700 exoplanets have been confirmed so far\footnote{\url{https://exoplanetarchive.ipac.caltech.edu/}} and the search continues. Detailed studies of planet hosts plays a key role in coming missions (e.g. PLATO:~\citealt{Rauer14}; Ariel:~\citealt{Tinetti21}) and advances the characterisation of exoplanets leading to a better understanding of their formation and evolution. Therefore, we have continued our high-resolution observation campaign to study bright stars (V$\leq$~8.5~mag) cooler than F5 spectral type by uniformly determining their main atmospheric parameters, ages, kinematic and orbital parameters, and elemental abundances. The results from our extensive study of 848 stars were published in a series of papers by \citet{ Tautvaisiene20,Tautvaisiene21,Tautvaisiene22} and another study on 249 dwarf stars were reported in papers by \cite{Mikolaitis18, Mikolaitis19, Stonkute20}. In this paper, we focus on the new sample of planet-hosting stars to analyse chemical elements (C, N, O, Mg, Si, and Fe) that are vital not only for stellar and Galactic chemical evolution, but also for exoplanet studies.

Theoretical studies on planet formation have demonstrated that relating the properties of exoplanets and the chemical composition of their host stars can offer a crucial explanation of the formation and evolution of these exoplanetary systems \citep[see e.g.,][]{Madhusudhan12, Unterborn19, Dorn19, Bitsch20, Mah23}. Well-established correlations with observations shows that stars with giant planets tend to have higher metallicities \citep[and references therein]{Gonzalez97, Santos01, Fischer05, Adibekyan19, Adibekyan21oct}. On the contrary, stars with low-mass planets do not seem to be preferentially metal-rich \citep{Ghezzi10, Sousa11, Buchhave12}. 

Such so-called volatile elements as carbon, nitrogen, and oxygen are key ingredients in stellar formation and evolution; their abundances also have a significant impact on planet formation \citep{Bitsch20,Bitsch23}. The C, N, and O abundances and their ratios (C/O; N/O) in exoplanet atmospheres have been used as probes to explore whether a given planet formed within or beyond the ''snow lines'' of various carbon, nitrogen, and oxygen-bearing molecules \citep{Oberg11, Schneider21, Ohno23}. The refractory elements are also relevant in the planet formation studies \citep{Chachan23}. The magnesium-to-silicon (Mg/Si) ratio governs the distribution of silicates in the planets \citep{Bond10}. The work by \cite{Thiabaud15} showed that this ratio in planets is essentially identical to those in the host stars.

The role of elemental abundances in the planet-host atmospheres has been studied using high-resolution spectra. For example, the C, O elemental abundances were determined in the works of \cite{Ecuvillon04b, Ecuvillon06, DelgadoMena10, Nissen14, Suarez-Andres17, Suarez-Andres18, DelgadoMena21, Mishenina21, Tautvaisiene22, Unni22}; with the N abundance in \cite{Ecuvillon04a, Suarez-Andres16, Biazzo22} and the Mg and Si abundances in \cite{Gonzalez09,Adibekyan12,Tautvaisiene22}. However, there is still considerable uncertainty, especially in the abundances of elements determined in different sample sizes of planet-hosts studies (e.g. small samples could potentially lead to uncertain conclusions). Also, the detailed abundances of CNO elements are difficult to determine for several reasons. The available atomic and molecular lines can present conflicting results \citep[see e.g.][]{Ecuvillon04a, Nissen14}. Moreover, at very low metallicities, the non-local thermodynamic equilibrium (NLTE) and 3D effects can have a substantial impact on the abundance measurements \citep{Amarsi22}. 
    
Following the discovery of a significant sample of exoplanets, observational studies have attempted to assess the importance of various elements by analysing the star-planet connection. For example, in the work by \cite{Ecuvillon04b}, carbon abundance in a set of planet-harbouring stars was determined from atomic carbon lines and results showed no appreciable difference in [C/Fe] between stars with and without planets. Other works studied carbon in a larger set of solar-type stars from the CH band at 4300~\AA~ and concluded that planet hosts are carbon-rich when compared to single stars \citep{Suarez-Andres17}. These authors also looked for a relationship between the carbon abundance and planetary masses, but all of the planetary masses showed a flat trend, indicating the absence of a significant contribution of carbon towards the mass of planets. More recently, tentative evidence was found that stars hosting low-mass planets have [C/Fe] higher than their counterparts without planets for [Fe/H]$<-0.20$. For more metal-rich hosts, no carbon enhancement associated with the presence of exoplanets was found \citep{DelgadoMena21}. 
    
Recent studies on oxygen abundance in planet-hosts suggest some hints of similar trends to those observed for carbon in stars with low-mass planets. However, due to the larger errors and small sample sizes in these studies affecting the results, no strong conclusions were drawn \citep{DelgadoMena21, Biazzo22}.

Nitrogen abundance is the least studied chemical element in stars with planets. The work by \citep{Suarez-Andres16} analysed 42 solar-type planet-hosts and derived nitrogen based on a spectral synthesis of the near-UV NH band at 3360~\AA. Their analysis showed that the number of stars with derived nitrogen abundance is not statistically significant to confirm the increase of nitrogen abundance for low-mass planets, followed by constant [N/Fe] value for the more massive planets. 

Looking at other chemical species that are important in the context of the mineralogy of planetary companions, an overabundance of alpha-elements (e.g. Mg, Si) in stars with planets was advocated \citep[e.g.][]{Adibekyan12, Tautvaisiene22}, especially in thick-disc stars. These findings might suggest that alpha-elements can account for lower Fe content during planet-building block formation \citep[e.g. see][]{Bashi19}.

A precise and homogeneous determination of the stellar atmospheric parameters and chemical compositions of planet-hosts in different Galactic components are mandatory to improve our knowledge of exoplanet formation and evolution. Homogeneous abundances of a statistically significant number of planet-hosts and a large enough comparison sample of stars are needed to reach this goal. Therefore, we are running a follow-up programme of planet-hosts with the high-resolution Vilnius University Echelle Spectrograph \citep{Jurgenson16} at the Moletai Observatory's 1.65 m telescope. Our comparison sample of bright stars without detected planets (1071 stars) are taken from our previous works \citep{Mikolaitis19, Stonkute20, Tautvaisiene22}, where the atmospheric parameters and elemental abundances are derived using the same methods as in this paper.  

This work is organised as follows. In Sect.~\ref{sec:data}, we describe the observations, and data reduction and present the methodology to determine stellar parameters together with chemical abundances. In Sect.~\ref{sec:results}, we present the results and discussion of our analysis. Finally, in Sect.~\ref{sec:conclusions}, we highlight our main conclusions.

\section{Observational data and analysis} \label{sec:data}

   \begin{figure*}
    \centering
    \includegraphics[width=0.9\hsize]{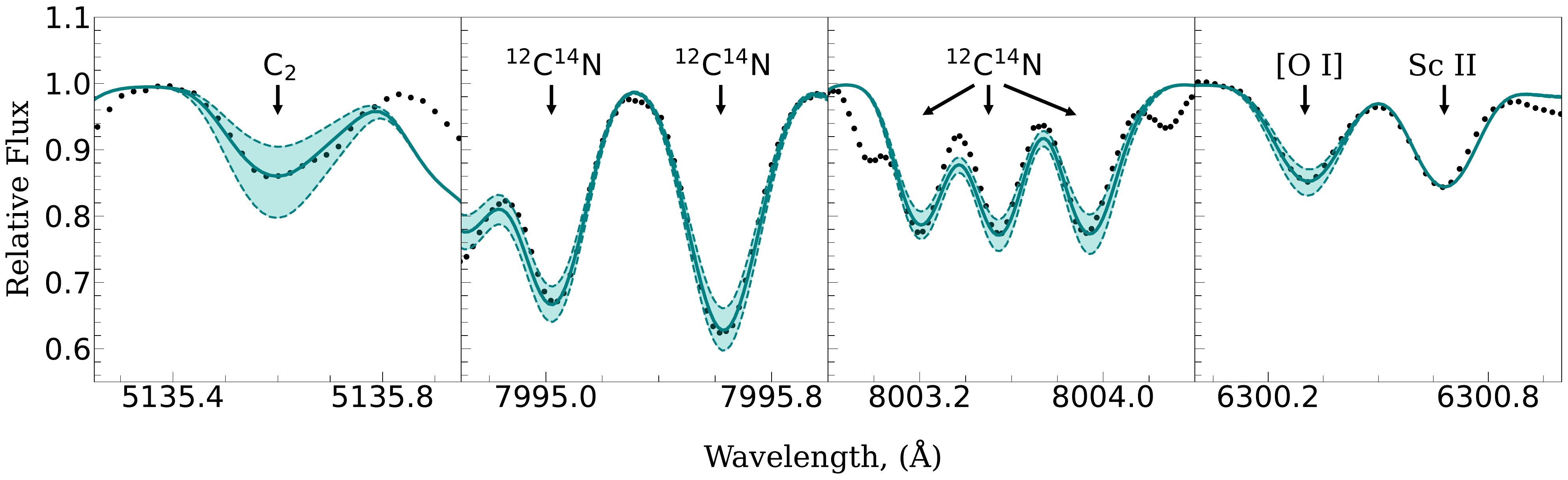}
    \caption{Example of spectral synthesis method showcasing the fitting of prominent spectral features including molecular carbon (${\rm C}_2$) band head located at 5135~{\AA}, molecular CN bands at 7995 and 8003~{\AA}, alongside the atomic forbidden oxygen line [O{\sc i}] at 6300~{\AA}. The observed spectra are represented by black dots, while the solid green lines denote the optimal fit of synthetic spectra to the observed data along with variations of $\pm$0.10~dex from the optimal fitted abundance.}
    \label{fig:spectrumfits}
   \end{figure*}
   
\subsection{Target selection and data reduction} \label{subsec:targets}
We selected bright (V$\leq$8.5 mag) stars from the NASA Exoplanet Archive and Transiting Exoplanet Survey Satellite list of confirmed and candidate exoplanets. The spectra of these planet hosts were observed with the high-resolution Vilnius University Echelle Spectrograph (VUES), installed on a 1.65 m telescope set up at Vilnius University's Moletai Astronomical Observatory in Lithuania \citep{Jurgenson16}. The spectrograph has $\sim$36,000, $\sim$51,000, and $\sim$68,000 resolution modes and covers the wavelength range from 400 to 880 nm. For our work, we used two resolution modes: R$\sim$36,000 \& 68,000. Depending on the stellar magnitudes, our observations had exposure time ranging from 900 to 7200 seconds and signal-to-noise ratios (S/Ns) ranging from 75 to 200. All the observations were carried out from 2021 through 2024. The data reduction was carried out on-site using an automated pipeline described in the work of \citet{Jurgenson16}.

\subsection{Stellar atmospheric parameters and chemical element abundances} \label{subsec:parametersandabundances}
We uniformly determined the main atmospheric parameters (effective temperature, $T_{\rm eff}$; surface gravity, ${\rm log}~g$; microturbulence velocity, $v_{\rm t}$; and metallicity ${\rm [Fe/H]}$) using the classical equivalent width approach. We used a combination of the DAOSPEC \citep{Stetson08} and MOOG programme codes \citep{Sneden73} to measure the equivalent widths (EWs) of atomic neutral and ionized iron lines (\ion{Fe}{i} and \ion{Fe}{ii}, respectively) to determine the stellar atmospheric parameters, in the same way, the Vilnius node used in the \textit{Gaia}-ESO Survey (\citealt{Smiljanic14} and \citealt{Mikolaitis18}). Effective temperature was derived by minimising the slope of iron abundances, obtained from \ion{Fe}{i} lines with respect to the excitation potential. Surface gravity was determined by enforcing ionization equilibrium, ensuring that the derived iron abundances from \ion{Fe}{i} and \ion{Fe}{ii} lines were the same. Microturbulence velocity was obtained by requiring that \ion{Fe}{i} abundances show no correlation with respect to the EWs of the lines.

After determining the stellar atmospheric parameters, we set out to determine the precise chemical abundances of carbon, oxygen, nitrogen, magnesium and silicon, using the spectral synthesis method with the TURBOSPECTRUM code \citep{Alvarez98}. The calculations were performed using a grid of MARCS stellar atmosphere models \citep{Gustafsson08}. For the models, we adopted the solar abundance mixture from \cite{Grevesse07} as the reference. To maintain consistency with our previous studies and the \textit{Gaia}-ESO Survey, we continue to use the same solar abundances from \cite{Grevesse07}. This approach ensures the uniformity and comparability of our results with our prior analyses.
      
Two molecular lines were selected for the carbon abundance determination. One is the ${\rm C}_2$ Swan (1, 0) band head at 5135~{\AA} and the other is ${\rm C}_2$ Swan (0, 1) band head at 5635~{\AA} \citep{Brooke13, Ram14}. The $\mathrm{^{12}C^{14}N}$ molecular lines in the regions 6470--6485~\AA~and 7980--8005~\AA~were used \citep{Sneden14} to derive nitrogen abundance. The forbidden [O\,{\sc i}] line at 6300~{\AA} was used to determine the abundance of oxygen. 
   
The C, N, and O chemical elements are constrained by the molecular equilibrium so these elements require a more specific approach. The correct determination of carbon (or oxygen) abundance can be carried out only if the value of the other is known, since, for cooler stars, much of carbon and oxygen are in the CO molecular form. The same applies to the nitrogen determination, as much of it is in CN molecules. Thus, the whole process involves multiple iterations of C and O abundances until the determinations of both of them converge. Subsequently, we enter the determined carbon and oxygen abundances when determining nitrogen as well. We refer to the example of the spectral synthesis method showcasing the fitting of prominent spectral features in Fig.~\ref{fig:spectrumfits}.
The non-local thermodynamical equilibrium (NLTE) effects on abundances of molecular carbon (${\rm C}_2$) and molecular nitrogen (CN) are expected to be negligible \citep{Ayres89}. In the work by \cite{Ryabchikova22}, it was argued that NLTE effects should not affect the abundances significantly since the atomic and molecular carbon and nitrogen give the same abundances for solar-type stars. On the other hand, the forbidden oxygen line at 6300~\AA\ has been studied and it was concluded that this line can be described in LTE \citep{Amarsi21}. 
   
For the \ion{Mg}{i} abundance determination, we used up to four (5528, 5711, 6318, and 6319 \AA) and for \ion{Si}{i} up to 12 spectral lines. The chemical abundances of \ion{Mg}{i} were determined with local thermodynamical equilibrium (LTE) and then we checked the influence of NLTE effects, as described in \cite{Bergemann17}. The corrections did not exceed 0.04~dex for \ion{Mg}{i}~line at 5711~\AA, while the other Mg lines need less than 0.01 dex corrections on average. In most of the cases, the strong \ion{Mg}{i}~line at 5711.07~\AA~was unused for the average Mg abundance determination.

\subsection{Determination of uncertainties} 
\label{subsec:errors}

   \begin{figure*}
    \centering
    \includegraphics[width=\hsize]{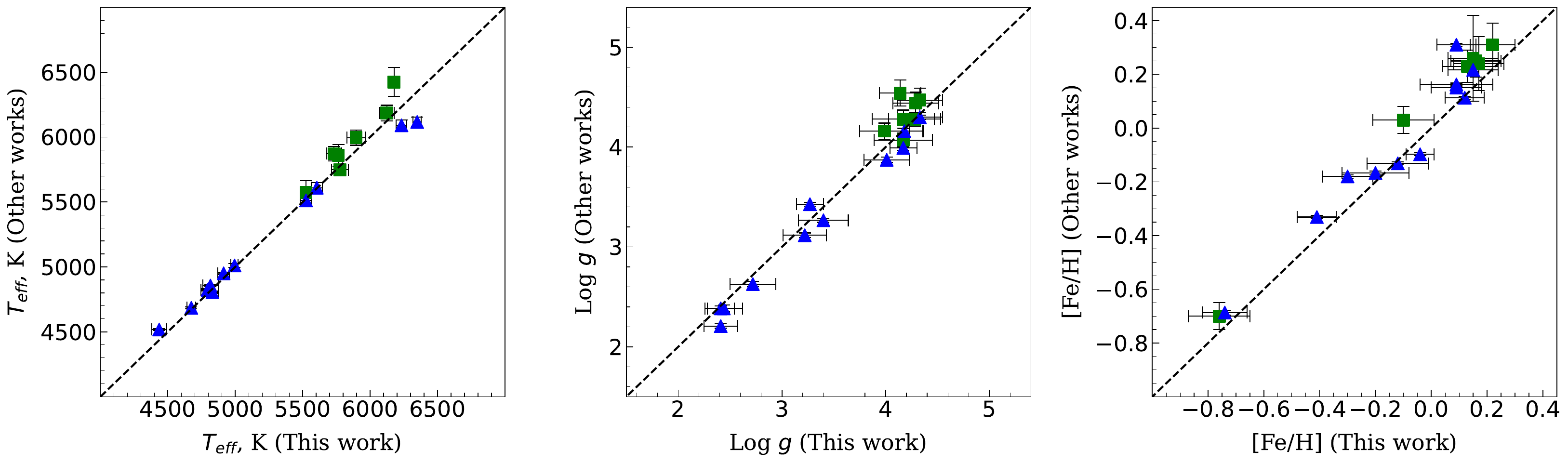}
    \caption{Spectroscopic stellar parameters ($T_{\rm eff}$ (left panel); ${\rm log}~g$ (middle panel); and ${\rm [Fe/H]}$ (right panel)) derived in this study compared with those reported by \citet{Bensby14}, represented by green squares, and APOGEE DR17 \citep{Abdurrouf22}, represented by blue triangles.}
    \label{fig:comparison}
   \end{figure*}

   \begin{table}
    \caption{Effect of uncertainties in atmospheric parameters on the derived chemical abundances for the target stars.} 
    \label{table:effects}      
    \centering
    \renewcommand{\arraystretch}{1.35}
    \begin{tabular}{l c c c c}
    \hline
    \hline
    {\raisebox{-1.5ex}[0cm][0cm]{Elements}}  &  $\Delta T_{\rm eff}$  &  $\Delta {\rm log}~g$  &  $\Delta {\rm [Fe/H]}$  &  $\Delta v_{\rm t}$ \\
    &  $\pm$ 50\,\text{K}  &  $\pm$ 0.20\,\text{dex}  &  $\pm$ 0.09\,\text{dex}  &  $\pm$ 0.25\,\text{km/s} \\
    \hline
    Dwarfs\\
    \hline
    C (C$_{2}$)          &  $\pm$ 0.03  &  $\mp$ 0.01  &  $\pm$ 0.02  &  $\pm$ 0.00 \\  
    N (CN)               &  $\pm$ 0.08  &  $\mp$ 0.01  &  $\pm$ 0.02  &  $\pm$ 0.00 \\
    O [O\,{\sc i}]       &  $\pm$ 0.02  &  $\pm$ 0.10  &  $\mp$ 0.09  &  $\pm$ 0.00 \\      
    Mg                   &  $\pm$ 0.03  &  $\mp$ 0.02  &  $\mp$ 0.01  &  $\mp$ 0.02 \\
    Si                   &  $\pm$ 0.01  &  $\pm$ 0.02  &  $\pm$ 0.00  &  $\pm$ 0.02 \\
    \hline
    Giants\\
    \hline
    C (C$_{2}$)          &  $\pm$ 0.00  &  $\pm$ 0.03  &  $\pm$ 0.02  &  $\pm$ 0.00  \\  
    N (CN)               &  $\pm$ 0.00  &  $\pm$ 0.05  &  $\pm$ 0.02  &  $\pm$ 0.00  \\
    O [O\,{\sc i}]       &  $\pm$ 0.00  &  $\pm$ 0.09  &  $\pm$ 0.01  &  $\mp$ 0.00  \\      
    Mg                   &  $\pm$ 0.02  &  $\pm$ 0.01  &  $\pm$ 0.00  &  $\mp$ 0.02  \\
    Si                   &  $\mp$ 0.00  &  $\pm$ 0.04  &  $\pm$ 0.00  &  $\mp$ 0.04  \\
    \hline
    \end{tabular} 
   \end{table}

We have encountered various potential sources of uncertainties in this work that were estimated at each step of the analysis. As mentioned before, we used the equivalent width method for the determination of stellar atmospheric parameters. In this step, we encounter uncertainties that can be attributed to the measurement of the lines (fitting of the lines, continuum placement etc.) or the method used to determine the parameters once we have the measurements (e.g. the linear regression fits). 

To mitigate the effect of the line measurement uncertainties, we used an extensive list of carefully selected \ion{Fe}{i} and \ion{Fe}{ii} lines (86 and 7, respectively), which were selected to avoid any contamination by blends, telluric lines, or to avoid regions with difficult continuum determination. When evaluating the uncertainties for the determined atmospheric parameters, we used the standard deviation of linear regression fits and abundances. Those deviations were propagated to find the boundary conditions for effective temperature, surface gravity and microturbulent velocity.

The typical uncertainties in our measurements are as follows: the average uncertainty in effective temperature, $T_{\rm eff}$, is approximately $50\pm15$~K, in surface gravity, ${\rm log}~g$, is $0.20\pm0.05$~dex, in metallicity, [Fe/H], is $0.09\pm0.02$~dex and in microturbulent velocity, $v_{\rm t}$,  is $0.25\pm0.08$~km/s. The uncertainty values are presented in Table~\ref{table:Results}.
  
Figure \ref{fig:comparison} shows a comparison of the spectroscopic stellar parameters derived in this work with the results of other works. We identified seven stars in common with \citet{Bensby14} work and eleven stars common with the APOGEE DR17 \citep{Abdurrouf22}. The comparison shows that the derived values are consistent with those from the literature, within the expected uncertainties.

We calculated any changes in abundances caused by the error of each atmospheric parameter, keeping other parameters fixed, as presented in Table~\ref{table:effects}. The results indicate that the abundances are not very sensitive to the changes in atmospheric parameters, both for dwarfs and giants. We see that only oxygen abundance is sensitive to the surface gravity.

\subsection{Kinematics and orbital properties} \label{subsec:kinematics}

   \begin{figure}
    \centering
    \includegraphics[width=\hsize]{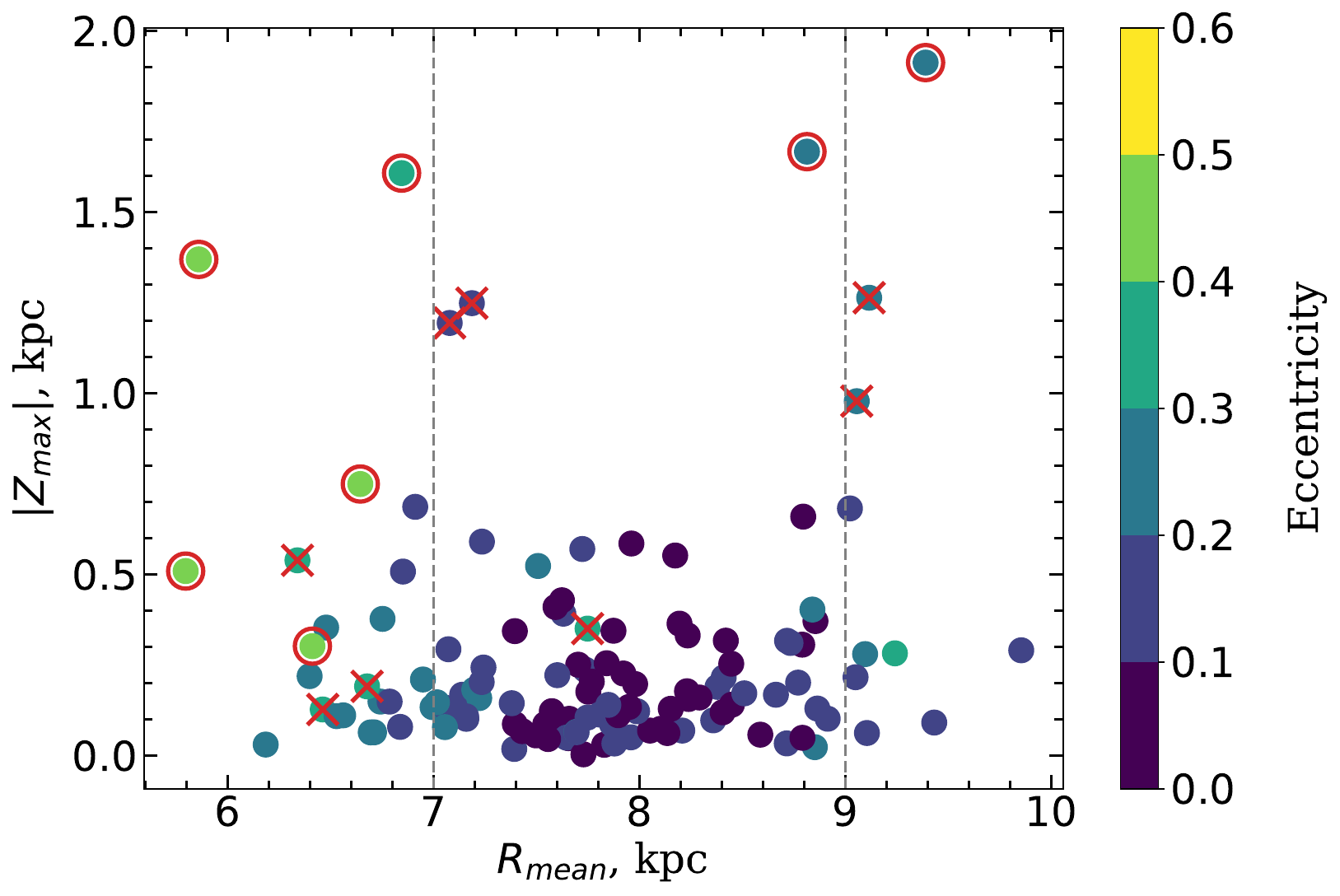}
    \caption{Mean Galactocentric distance ($R_{\rm mean}$) as a function of the maximum Galactic height, $|z_{\rm max}|$ of observed stars. Stars are colour-coded by their orbital eccentricity. The two vertical dashed lines indicate the Solar neighbourhood 7$<R_{\rm mean}<$9~kpc. The data points with outer red circles represent thick disc stars while those with crosses represent stars in between the thin and thick discs.}
    \label{fig:RmeanvsZmax}
   \end{figure}

  \begin{figure}
    \centering
    \includegraphics[width=\hsize]{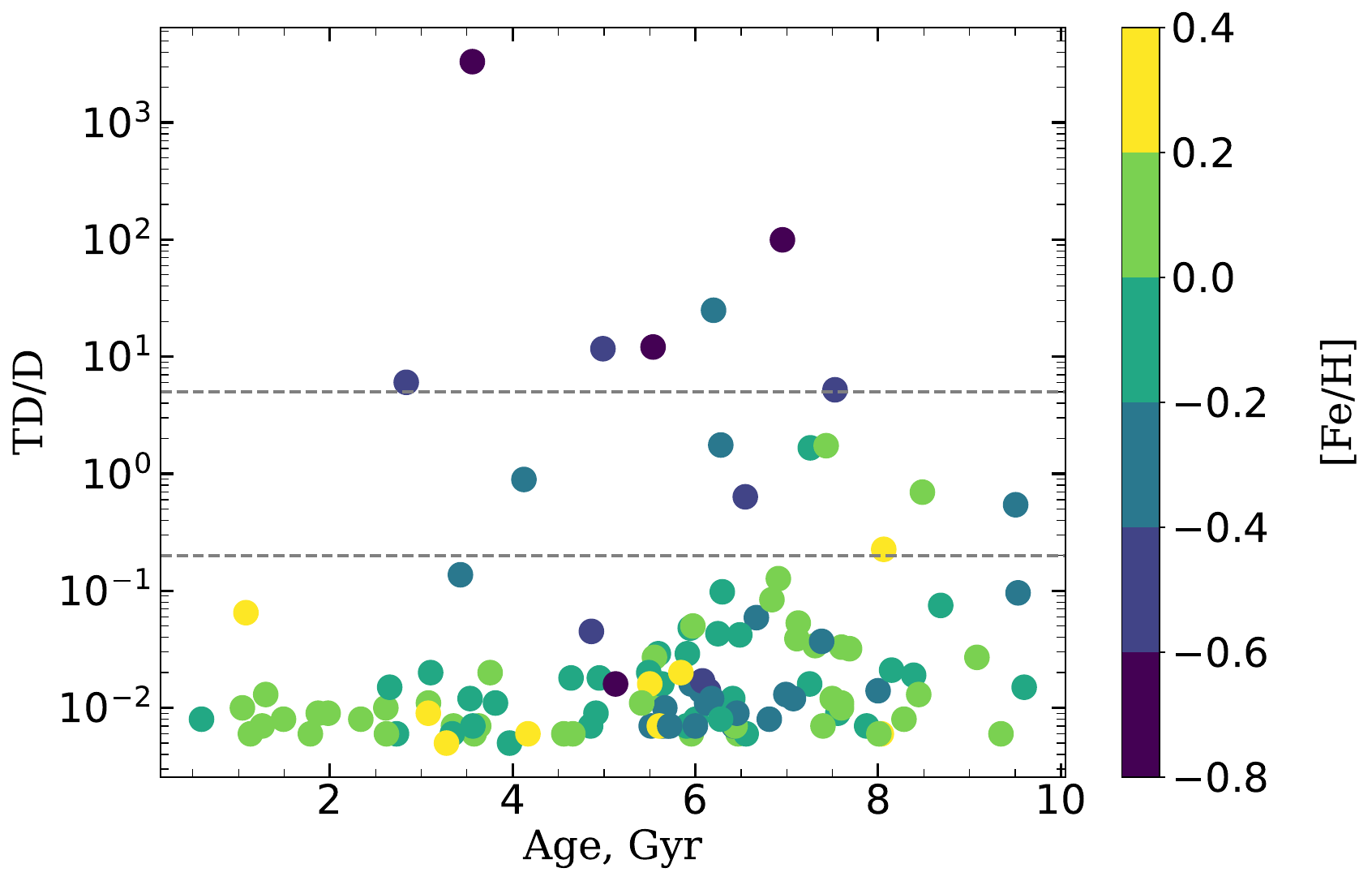}
    \caption{Age versus kinematical thick disc-to-thin disc probability ratio (TD/D) colour coded by metallicity, [Fe/H].}
    \label{fig:AgevsTD_D}
   \end{figure}

   \begin{figure}
    \centering
    \includegraphics[width=\hsize]{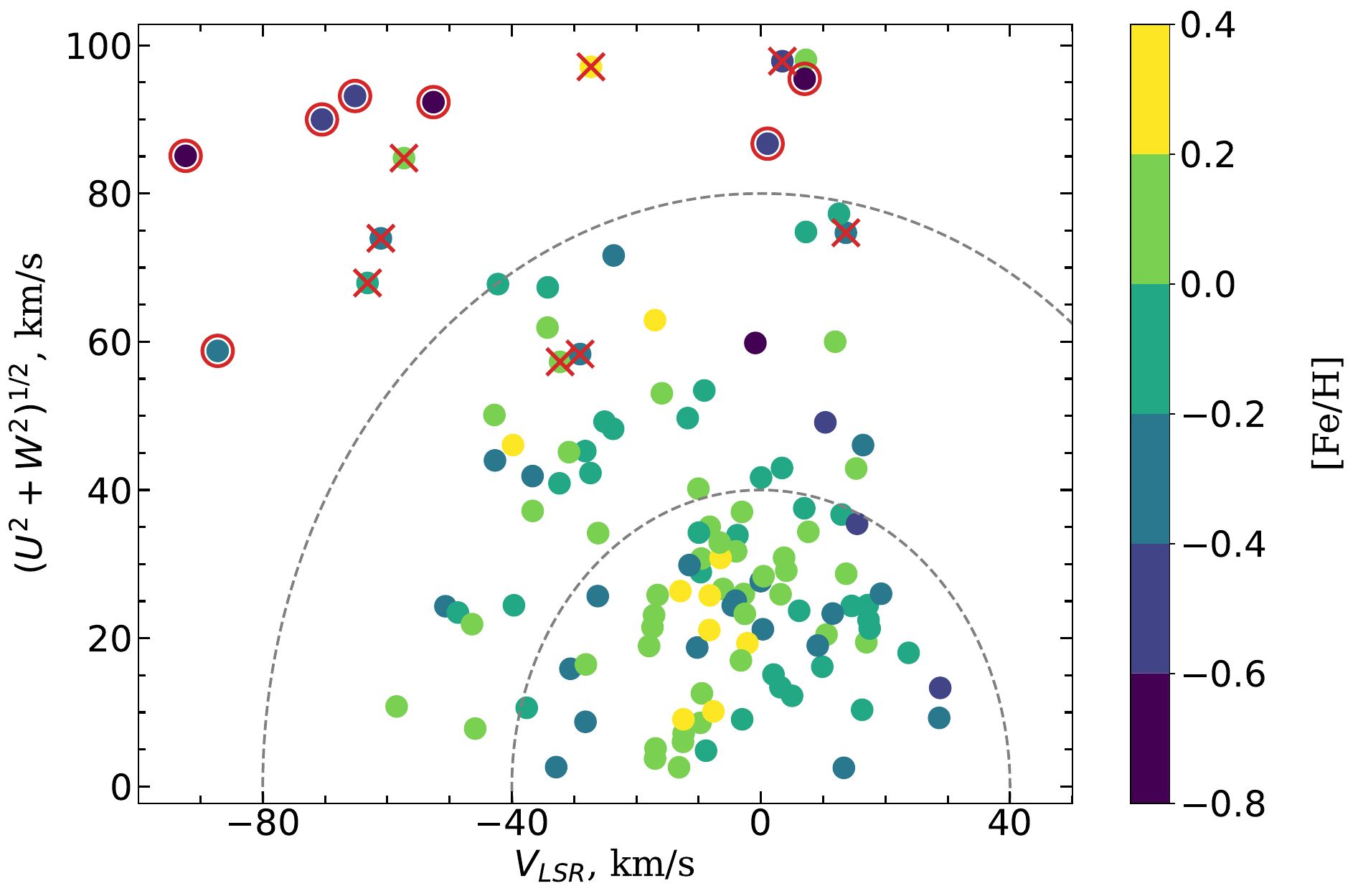}
    \caption{A Toomre diagram for the stars, where the symbols have been coded according to the stellar metallicity, [Fe/H]. The dashed lines represent the constant total space velocity ($v_{\rm tot}=(U_{\rm LSR}^2+V_{\rm LSR}^2+W_{\rm LSR}^2)^{\rm 1/2)}$) values at 40 and 80~km\,s$^{\rm -1}$. The data points with outer red circles represent thick disc stars while those with crosses represent stars in between the thin and thick discs.}
    \label{fig:Toomre}
   \end{figure}

The dynamical history of stars, including stellar kinematics, orbital properties, and stellar ages, may impact the planets' distribution and architecture in the Galaxy \citep{Adibekyan21may}. Therefore, we aim to study our planet-host sample by taking into account their kinematics and orbital properties. 

We derived the Galactic space velocities (\textit{U}, \textit{V}, \textit{W}), mean galactocentric distance, $R\mathrm{_{mean}}$,  maximum vertical distance from the Galactic plane, $|z_{\rm max}|$, and the eccentricity, $e$. To derive these values, we used the Python-based package for galactic-dynamics calculations \textit{galpy} \footnote{\url{http://github.com/jobovy/galpy}} by \cite{Bovy15}. To perform the calculations of galactic orbits, stellar distances were taken from \cite{BailerJones21}, other stellar parameters such as proper motions and stellar coordinates were taken from the \textit{Gaia} data release 3 (EDR3) catalogue \citep{GaiaCollaboration16, GaiaCollaboration21, Lindegren21, Seabroke21} and radial velocities for all our targets were determined through our own calculations. The input data of star "HD 62509" (V=1.14 mag) is not available in the \textit{Gaia} EDR3 catalogue and, thus, it was taken from the Hipparcos catalogue \citep{vanLeeuwen07}.

We defined a gravitational potential of the Milky Way Galaxy using the MilkyWayPotential2014 model and initialised an orbit integration scheme in \textit{galpy} to compute the trajectories of the stars in the defined potential. The \textit{galpy} was set to integrate orbits for 5~Gyr. One thousand Monte Carlo computations were performed to determine observational errors in the orbital properties based on errors in input parameters. We used the position and movements of the Sun from (\cite{Bovy12}; $R_{\rm gc\odot}=8$~kpc and $V_{\odot}=220$~km\,s$^{-1}$), the distance from the Galactic plane $z_{\odot}=0.02$~kpc \citep{Joshi07}, and the local standard of rest velocities (LSR) from (\cite{Schonrich10}; \textit{U, V, W} = 11.1, 12.24, 7.25~km\,s$^{-1}$).

The kinematic parameters, along with their corresponding standard deviation values, are detailed in Table~\ref{table:Results}. Figure \ref{fig:RmeanvsZmax} represents the distribution of stars in the $z\mathrm{_{max}}$ versus $R\mathrm{_{mean}}$ plane, colour-coded by eccentricity. The two vertical dashed lines on the plot serve as boundaries, demonstrating the region of the solar neighbourhood with 7$<R_{\rm mean}<~$9~kpc.

\subsection{Stellar age determinations} \label{subsec:ages}
For stellar age determinations, we used the unified tool to estimate distances, ages, and masses (UniDAM) \cite{Mints17, Mints18}. The code uses a Bayesian approach and the PARSEC isochrones \citep{Bressan12} together with stellar atmospheric parameters (we used our spectroscopically derived parameters) and infrared magnitudes. The required \textit{J}, \textit{H}, and \textit{K} magnitudes from the Two Micron All-Sky Survey (2MASS, \citealt{Skrutskie06}) and the $W1$ and $W2$ magnitudes from AllWISE \citep{Cutri14} were taken as input for UniDAM. 

UniDAM incorporates state-of-the-art stellar evolution models and data analysis algorithms. By comparing the observed properties of the stars with the theoretical isochrones, we were able to determine their ages. However, we need to    take into account that UniDAM assumes a scaled solar abundance pattern and this can introduce a bias in the age estimates when this assumption is wrong.
  
   \begin{figure}
    \centering
    \includegraphics[width=\hsize]{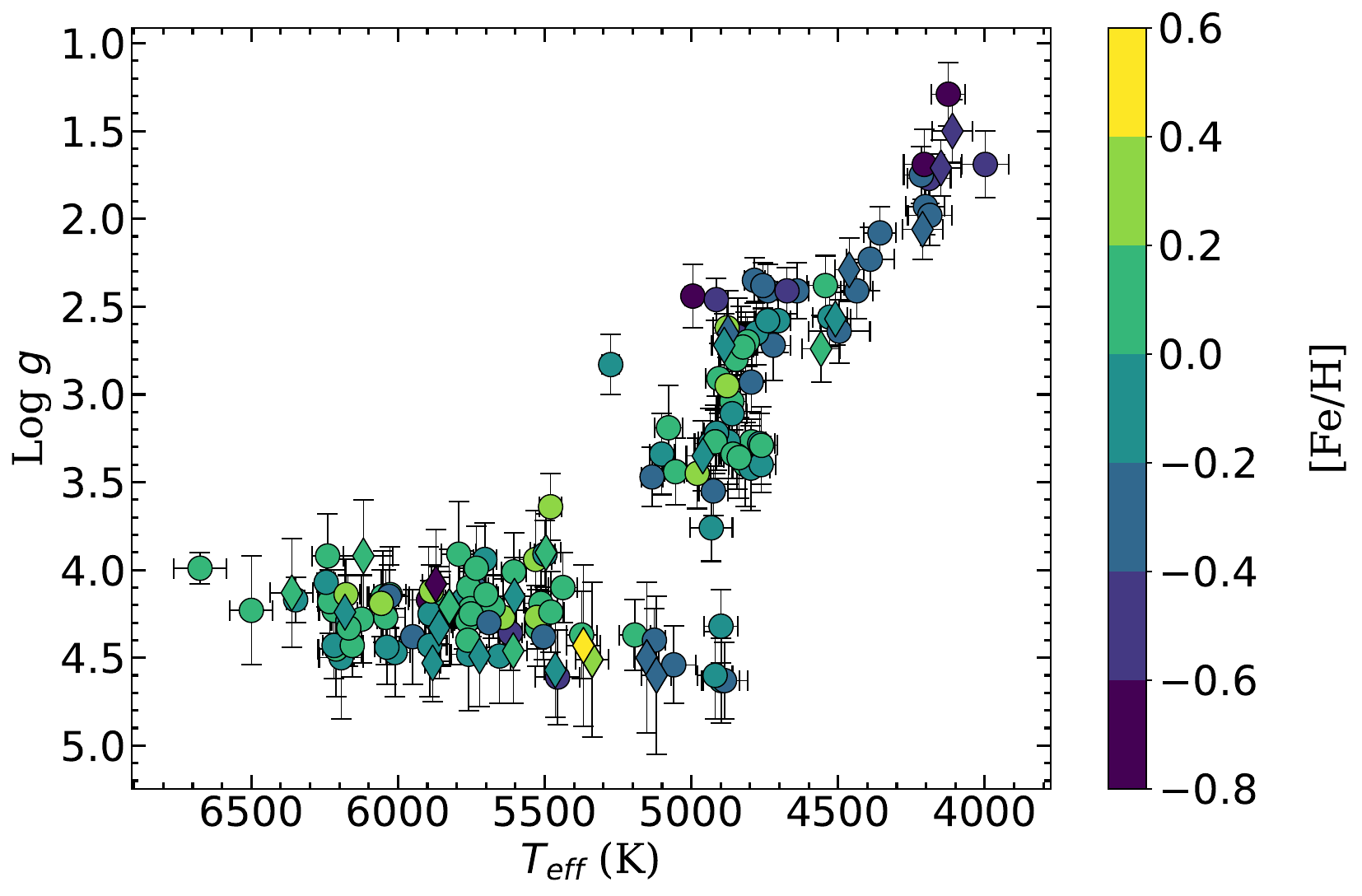}
    \caption{Effective temperature ($T_{\rm eff}$) versus surface gravity (${\rm log}~g$) diagram for observed planet host stars, colour-coded by metallicity, [Fe/H].}
    \label{fig:Teffvslogg}
   \end{figure}

   \begin{figure*}
    \centering
    \includegraphics[width=0.8\hsize]{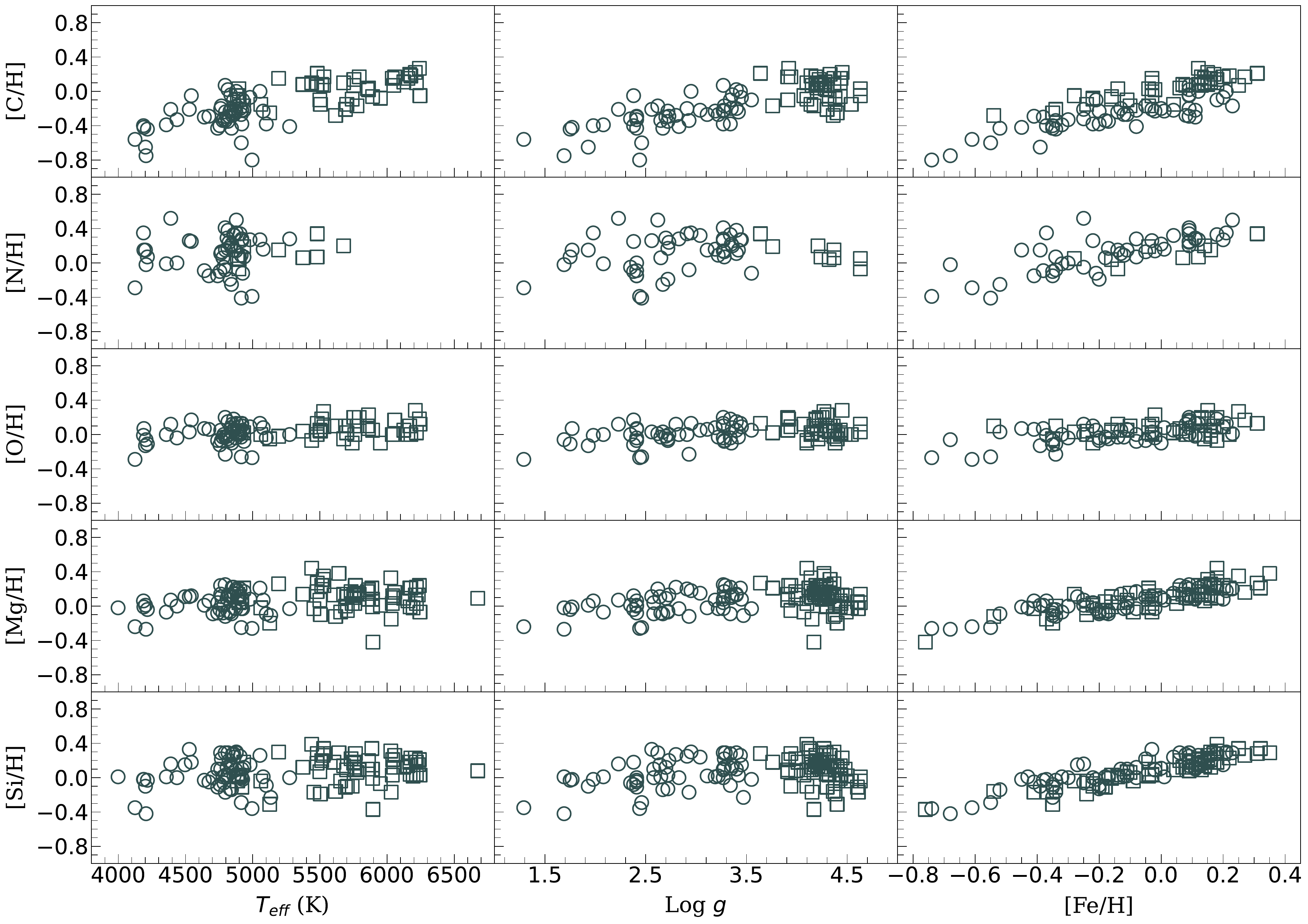}
    \caption{Derived C, N, O, Mg, and Si abundances relative to hydrogen versus the atmospheric parameters. Squares and circles represent dwarfs and giants, respectively.}
    \label{fig:XHvsParameters}
   \end{figure*}

\subsection{Kinematically selected thin and thick disc samples} \label{subsec:discratios}
Criteria used in this research for separating the stellar sample into Galactic discs involve an investigation of stellar kinematics and calculating the thick-to-thin disc probability ratios, (TD/D). This is a dimensionless ratio that expresses how much more likely it is for a particular star to belong to the thick disc, rather than the thin disc. Galactic thin disc and thick disc differ in properties in several aspects \citep{Bensby14}. The thin disc is generally younger and contains a higher proportion of young stars than the thick disc \citep{Gilmore85}. In Fig.~\ref{fig:AgevsTD_D}, we show the stellar age distribution with thin and thick disc-like kinematics, colour-coded by the host's [Fe/H]. Stars with TD/D$\geq$2 were attributed to the tick disc, while those with TD/D$\leq$0.5 were attributed to the thin disc. Other stars were assigned to the in-between sample.  We see that the metal-rich hosts show thin disc-like kinematics. 

In Fig.~\ref{fig:Toomre}, we plot a Toomre diagram that allows us to explore the kinematic properties of stars in a 2D space defined by their radial and vertical velocities. By examining the distribution of stars in the diagram, we can disentangle stars associated with the galactic thin and thick disc kinematics. The stars with a lower combined velocity $v_{\rm tot}$ \(<\) 50~km\,s$^{\rm -1}$ ($v_{\rm tot}=(U_{\rm LSR}+V_{\rm LSR}+W_{\rm LSR})^{\rm 1/2)}$) are most probably the thin-disc stars, whereas those with $v_{\rm tot}$ \(>\) 50 ~km\,s$^{\rm -1}$ should belong to the thick disc. As we can see in Fig.~\ref{fig:Toomre}, our kinematically selected thick disc stars (red circles) have total space velocity higher than 80~km\,s$^{-1}$. 

   \begin{figure*}
    \centering
    \includegraphics[width=0.9\hsize]{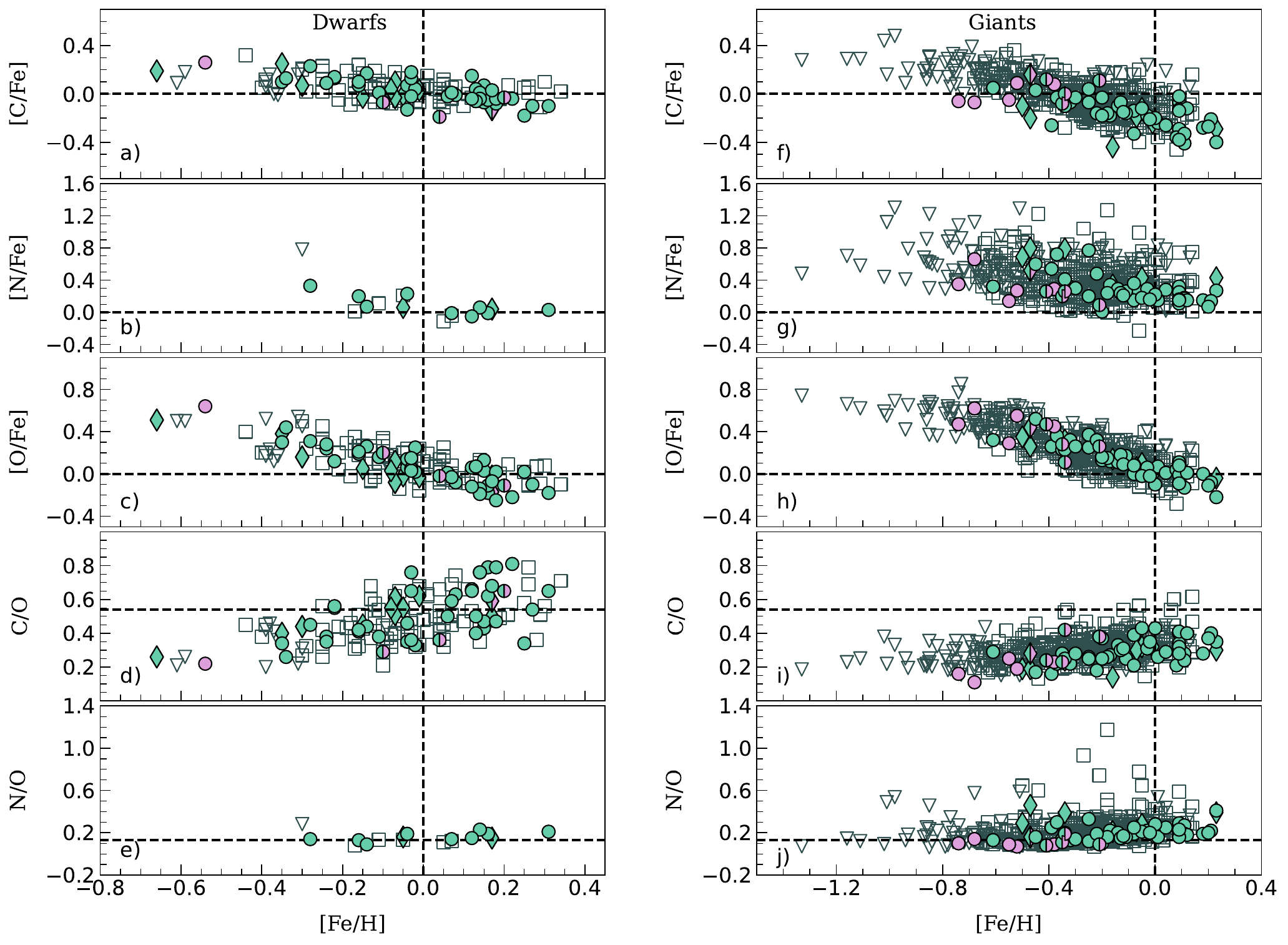}
    \caption{[C/Fe], [N/Fe], and [O/Fe] abundances and C/O and N/O ratios as functions of [Fe/H] for observed dwarfs (left panel) and giants (right panel). Thin disc planet-hosting stars are represented by green circles, while pink circles indicate stars from the thick disc. Circles filled with both colours represent "in between stars" to thin and thick discs. The (coloured) diamond symbols represent the planets-hosts from \cite{Tautvaisiene22}. The comparison sample results are taken from \cite{Stonkute20, Tautvaisiene22} and indicated by empty grey squares (thin disc) and triangles (thick disc).}
    \label{fig:CNOvsFeH}
   \end{figure*}

   \begin{figure}
    \centering
    \includegraphics[width=\hsize]{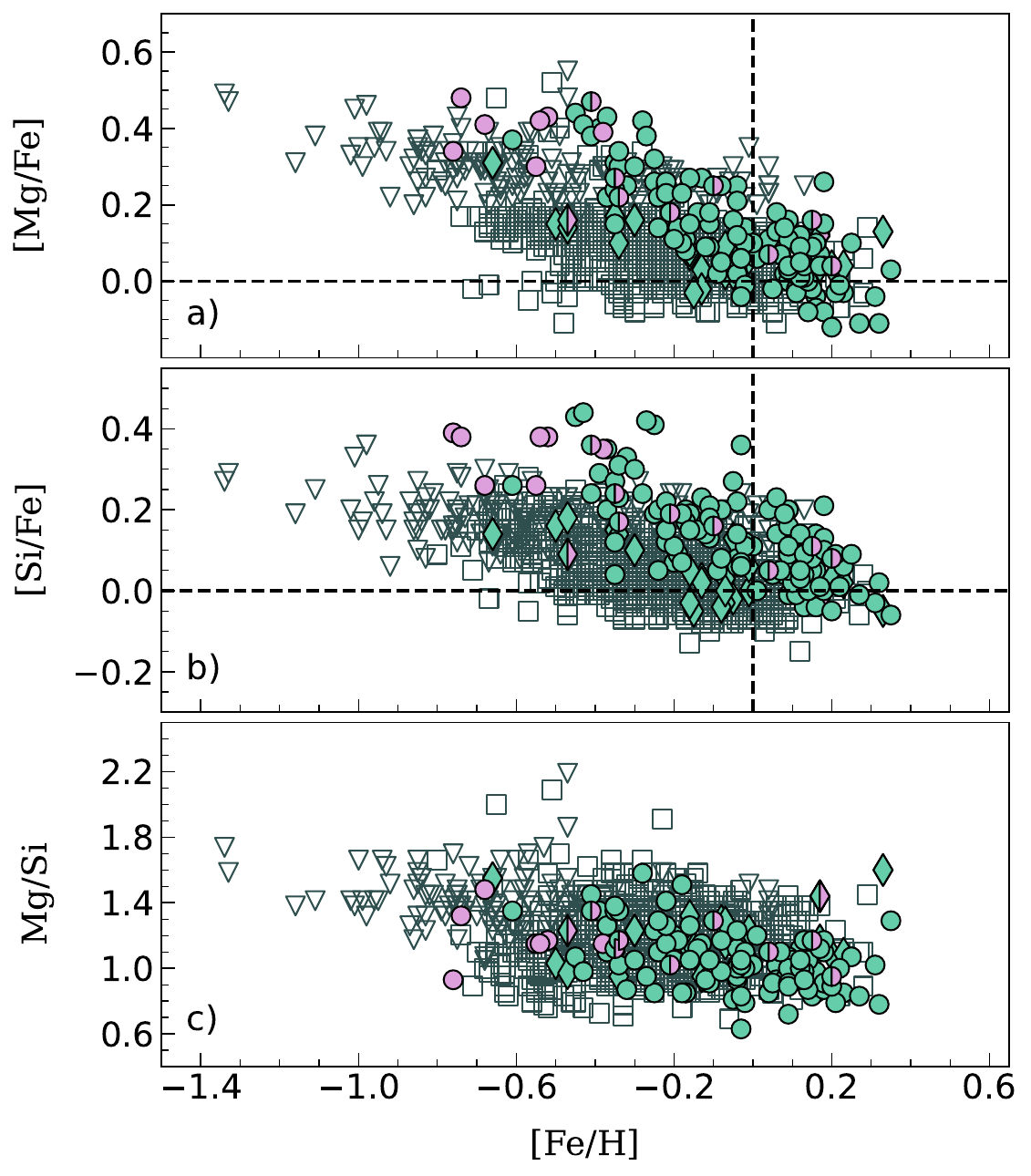}
    \caption{The [Mg/Fe], [Si/Fe], and Mg/Si ratio as functions of [Fe/H] for all investigated stars. All symbols have the same meaning as in Fig.~\ref{fig:CNOvsFeH}. The comparison sample results are taken from \cite{Mikolaitis19, Tautvaisiene22} and indicated by the empty grey squares (thin disc) and triangles (thick disc).}
    \label{fig:MgSivsFeH}
   \end{figure}

\section{Results and discussion}\label{sec:results}

\subsection{Stellar parameters} \label{sec:Results_Parameters} 
For this work, we compiled a newly observed sample of 124 stars with associated planets observed with the 1.65~m Moletai Observatory telescope. We expanded this sample to a final sample of 149 stars by adding 25 planet-hosts from our previous works \citep{Stonkute20,Tautvaisiene22} analysed using the same methodology as in this work. This 149-star sample consists of 83 main sequence stars referred in this paper as dwarf stars, along with 66 stars that are at their evolved stages, referred in this paper as giant stars.
    
In Fig.~\ref{fig:Teffvslogg}, we presented the HR diagram of observed planet-hosts, colour-coded by metallicity. The effective temperature, $T_{\rm eff}$, of the observed stars, exhibits a broad range from 3998 to 6675~K. Surface gravity, ${\rm log}~g$, which highlights the intrinsic differences between dwarf and giant stars, ${\rm log}~g$, ranges from 1.3~to~3.5~dex for giant stars, with a mean value of $2.8\pm0.6$~dex, while dwarf stars exhibit ${\rm log}~g$ values ranging from 3.6~to~4.6~dex, with a mean value of $4.2\pm0.2$~dex. Metallicity [Fe/H], a key parameter indicative of the planet occurrence rate, ranges from $-0.76$ to 0.35~dex, with a mean of $-$$0.08\pm0.20$~dex. The atmospheric parameters for the stars under investigation are presented in Table~\ref{table:Results}. 
   
After determining the atmospheric parameters, we conducted a comprehensive analysis of the abundances of carbon, nitrogen, oxygen, magnesium, and silicon. To investigate any systematic differences in derived abundances, we present the results in Fig.~\ref{fig:XHvsParameters} as [X/H] versus $T_{\rm eff}$, ${\rm log}~g$, and [Fe/H]. Here, dwarfs and giants are represented by squares and circles, respectively. The results indicate that there are no systematic differences in the derived abundances, except for the expected stellar evolutionary effects, visible in the subplots of carbon and nitrogen. This is addressed in later sections of the text.

\subsection{Correlation between element abundances [X/Fe] and metallicity [Fe/H]} \label{subsec:Abundance_vs[Fe/H]} 
In Fig. \ref{fig:CNOvsFeH}, we depict the abundances of carbon, nitrogen, and oxygen along with their ratios (C/O and N/O) as functions of metallicity separately for dwarf and giant stars. Thin-disc planet hosts are denoted by green markers, thick-disc hosts by pink circles, and stars with kinematics between thin and thick discs are represented by circles filled with both colours, which we refer to as "in-between stars." The dashed black lines represent solar values. The comparison stars are sourced from our previous studies \citep{Stonkute20, Tautvaisiene22}, where the parameters and abundance determinations follow the same methodology. 

The majority of our sample stars have thin disc kinematics and our abundance analysis results align with the Galactic chemical evolution.  At lower metallicities, massive stars produce some carbon and predominantly oxygen, whereas iron comes from Ia supernova on longer time scales. At the same time, carbon abundance follows iron more closely than oxygen, because C is also produced in stars of all masses. Thus, the ejection of carbon abundance is delayed in time concerning oxygen, so the C/O ratio could potentially offer a relative age clock for populations of stars. At solar- and super-solar metallicities, the C and O abundances continue to decrease with increasing metallicity for planet hosts. The metal-rich dwarf hosts, on average, exhibit slightly lower C and O distributions than comparison stars. Our results show that 100\% of stars with planets have C/O values lower than 0.81.

The recent work by \cite{Unni22} analysed carbon (CH in the G-band region) derived from the LAMOST spectra in the Kepler field dwarf stars and found that there is a preference for giant planets around host stars with a subsolar [C/Fe] ratio and higher [Fe/H]. However, at lower metallicities (where mostly low-mass planet hosts are present), the planet hosts may have slightly higher [C/Fe] values than the field stars, which is similar to what is observed in $\alpha$-elements \citep{Adibekyan12}. 

Our results on C and O abundances and the C/O ratio for dwarf stars are also in agreement, within uncertainties, with the previous studies \citep{Suarez-Andres17, Suarez-Andres18, Stonkute20, Unni22} at a given metallicity range. However, the different trend in carbon abundance is seen at the super-solar metallicity end between our study and \citet{Suarez-Andres17}. This difference can be attributed to different carbon lines used to derive the abundance among the other possibilities (e.g. methods used to derive the C abundance). 

For metal-rich dwarf stars, the nitrogen abundance follows closely the solar value, however, we see a slight increase in the N abundance at sub-solar metallicity. Looking at the elemental N/O ratio, we see a hint of abundance increase with [Fe/H]. The small number of dwarf stars with planets and derived nitrogen abundances limit our ability to interpret these data with high confidence. Nevertheless, from the previous work by \citet{Ecuvillon04a}, we see that planet host and comparison sample stars have the same [N/Fe] versus [Fe/H] trends. Moreover, \citet{Suarez-Andres16} demonstrated that planet hosts are nitrogen-rich compared to single stars, however, considering the linear trend between [N/Fe] and [Fe/H], this can be explained by the metal-rich nature of planet hosts. We recognize the need for additional homogeneous nitrogen abundance data to shed light on nitrogen trends in planet hosts.

The C, N, and O abundance trends observed in giant stars are different than in dwarfs due to material mixing effects in evolved stars (e.g. \citealt{Lagarde19}). The carbon abundance is depleted by about 0.2~dex, nitrogen is enhanced by 0.2~dex, and the oxygen abundance is close to abundance in dwarfs at a given [Fe/H]. Consequently, the C/O ratio is lowered by about 0.25~dex and the N/O ratio is enhanced in giant stars. The trend between [C; N; O/Fe] and [Fe/H] in giant stars with comparison sample is similar; however, there is a presence of a smaller scatter in C and O abundances of planet hosts and on average lower carbon at a given metallicity. Furthermore, the C/O ratio in giant stars with planets seems to be on average lower as well. Also, looking at the elemental N/O ratio in giant stars, we find an increasing trend with [Fe/H], but the difference between the two samples is negligible (see Sect.~\ref{subsec:Statistics} for more discussion). 

Figure \ref{fig:MgSivsFeH} illustrates the distribution of the other two so-called $\alpha$-elements: magnesium and silicon. We show the [Mg/Fe] and [Si/Fe] as functions of [Fe/H] for both dwarf and giant stars determined in this work (see panel a) and b)). Additionally, in panel c), we show the Mg/Si ratio as a function of metallicity. Similar to the C, N, and O versus metallicity comparison, we included both planet hosts and a comparison sample taken from \cite{Mikolaitis19, Tautvaisiene22}. Each symbol has the same meaning as in Fig.~\ref{fig:CNOvsFeH}. Our results show that metal-poor stars exhibit higher Mg and Si abundances compared to metal-rich stars as expected from the Galactic chemical evolution. Also, we observe an overabundance of Mg and Si in planet hosts compared to the comparison sample, especially on the lower-metallicity side. This overabundance of $\alpha$-elements in stars with planets was demonstrated in works by \cite{Haywood08} and \cite{Adibekyan12} and indicates that $\alpha$-elements can play an important role in the formation of planets. The observed Mg/Si ratio appears to be on average slightly lower in stars with planets compared to the comparison sample at a given [Fe/H]. 

To comprehensively analyse any differences observed in abundance ratios C/O, N/O, and Mg/Si between the two stellar samples, we conducted two statistical tests: the Kolmogorov-Smirnov (K-S) and Anderson-Darling (A-D) tests. The outcomes of these tests are explained in detail in Sect. \ref{subsec:Statistics}.

\subsection{C/O, N/O, and Mg/Si in planet-hosts} 
\label{subsec:Statistics} 
    
    \begin{figure*}
    \centering
    \includegraphics[width=\hsize]{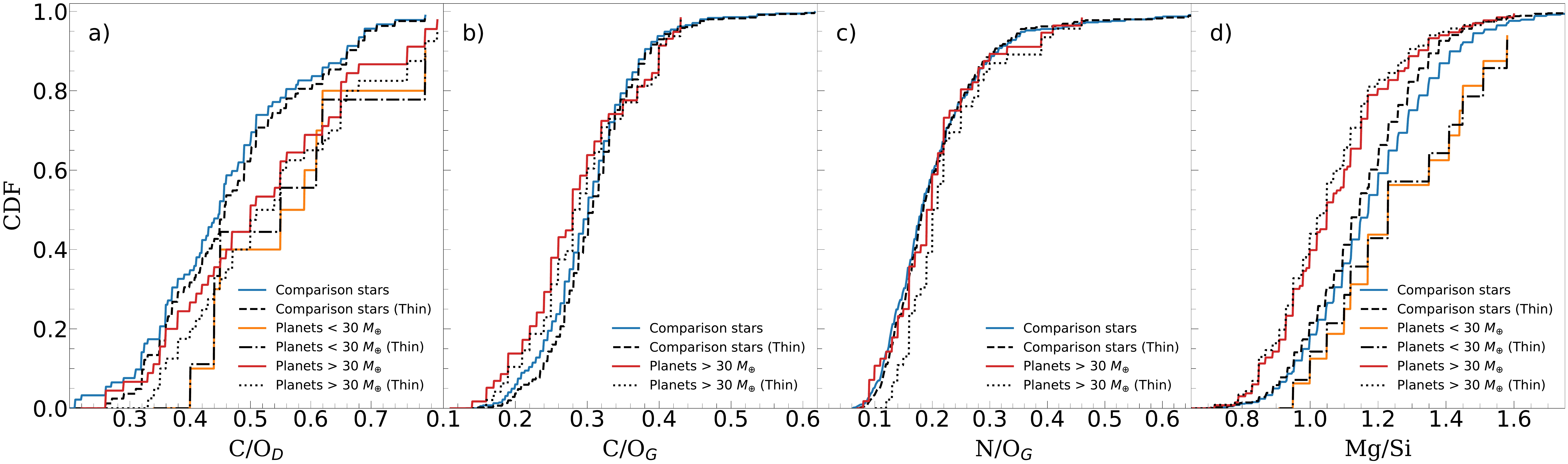}
    \caption{Cumulative distribution functions (CDFs) of C/O, N/O, and Mg/Si. In panel \textit{a:} Cumulative C/O$_{\text{D}}$ distributions for dwarf stars. Solid orange and red lines represent low-mass ($M_{\text{p}}<30M_{\text{$\oplus$}}$) and high-mass planet ($M_{\text{p}}>30M_{\text{$\oplus$}}$) hosts, while the solid blue line displays the comparison stars. Black lines represent the same as coloured, but only taking the stars with thin-disc kinematics. Panels \textit{b}, \textit{c:} Cumulative C/O$_{\text{G}}$; N/O$_{\text{G}}$ distributions for giant stars. Panel \textit{d:} Cumulative Mg/Si distributions for all stars with planetary companions and the comparison sample.}
    \label{fig:CDF}
   \end{figure*}
   
   \begin{table*}
    \caption{Kolmogorov-Smirnov (K-S) statistical test results for Mg/Si and C/O elemental ratio distributions for stars with and without planetary companions.} 
    \label{table:2}      
    \centering
    \renewcommand{\arraystretch}{1.35}
    \begin{tabular}{l c c c c c c c c}
    \hline
    \hline
    & \multicolumn{2}{c}{K-S$^{\text{Low mass}}_{\text{Whole sample}}$} & \multicolumn{2}{c}{K-S$^{\text{Low mass}}_{\text{Thin disc}}$} & \multicolumn{2}{c}{K-S$^{\text{High mass}}_{\text{Whole sample}}$} & \multicolumn{2}{c}{K-S$^{\text{High mass}}_{\text{Thin disc}}$} \\
    \cline{2-3} \cline{4-5} \cline{6-7} \cline{8-9}
    & Statistic & $p$-value & Statistic & $p$-value & Statistic & $p$-value & Statistic & $p$-value \\ 
    \hline
    C/O$_{\text{D}}$         & 0.28 & 0.39  & 0.21 & 0.81  & 0.23 & 0.07        & 0.21 & 0.17\\  
    C/O$_{\text{G}}$         &  --  &  --   &  --  &  --   & 0.22 & 0.01        & 0.22 & 0.02\\
    N/O$_{\text{G}}$         &  --  &  --   &  --  &  --   & 0.11 & 0.49        & 0.20 & 0.06\\      
    A(C+N+O)$_{\text{G}}$    &  --  &  --   &  --  &  --   & 0.14 & 0.24        & 0.23 & 0.02\\
    Mg/Si                    & 0.19 & 0.55  & 0.23 & 0.38  & 0.30 & $\leq{0.001}$ & 0.29 & $\leq{0.001}$\\
    \hline
    \hline
    \multicolumn{9}{l}{Two-sided K-S statistics test and p-values validated through 1000 bootstrap resampling iterations.}\\
    \hline
    C/O$_{\text{D}}$         & 0.38  & 0.20 & 0.34 & 0.34 & 0.25 & 0.13        & 0.24 & 0.20\\
    C/O$_{\text{G}}$         &   --  &  --  &  --  &  --  & 0.24 & 0.03        & 0.25 & 0.04\\
    N/O$_{\text{G}}$         &   --  &  --  &  --  &  --  & 0.16 & 0.21        & 0.22 & 0.09\\    
    A(C+N+O)$_{\text{G}}$    &   --  &  --  &  --  &  --  & 0.18 & 0.14        & 0.25 & 0.05\\
    Mg/Si                    & 0.27  & 0.30 & 0.29 & 0.27 & 0.31 & $\leq{0.001}$ & 0.31 & $\leq{0.001}$\\
    \hline
    \end{tabular}
    \tablefoot{The number of determined N abundances in dwarfs is small, so N/O and A(C+N+O) tests are inconclusive, and no results are presented.} 
   \end{table*}

   \begin{table*}
    \caption{Anderson-Darling (A-D) statistical test results for Mg/Si and C/O elemental ratio distributions for stars with and without planetary companions.}
    \label{table:3}
    \centering
    \renewcommand{\arraystretch}{1.35}
    \begin{tabular}{l c c c c c c c c}
    \hline
    \hline
    & \multicolumn{2}{c}{A-D$^{\text{Low mass}}_{\text{Whole sample}}$} & \multicolumn{2}{c}{A-D$^{\text{Low mass}}_{\text{Thin disc}}$} & \multicolumn{2}{c}{AD$^{\text{High mass}}_{\text{Whole sample}}$} & \multicolumn{2}{c}{A-D$^{\text{High mass}}_{\text{Thin disc}}$} \\
    \cline{2-3} \cline{4-5} \cline{6-7} \cline{8-9}
    & Statistic & $p$-value & Statistic & $p$-value & Statistic & $p$-value & Statistic & $p$-value \\     
    \hline
    C/O$_{\text{D}}$        & 0.42    &    0.22      & $-0.45$ & $\geq{0.25}$ & 1.71    &    0.06       & 1.07  & 0.12\\
    C/O$_{\text{G}}$        &  --     &      --      &   --    &     --       & 4.19    &    0.01       & 3.72  & 0.01\\
    N/O$_{\text{G}}$        &  --     &      --      &   --    &     --       & $-0.56$ & $\geq{0.25}$  & 1.20  & 0.10\\  
    A(C+N+O)$_{\text{G}}$   &  --     &      --      &   --    &     --       & 1.15    &    0.11       & 4.41  & 0.01\\
    Mg/Si                   & $-0.42$ & $\geq{0.25}$ &  0.32   & $\geq{0.25}$ & 29.82   & $\leq{0.001}$ & 24.87 & $\leq{0.001}$\\
    \hline
    \hline
    \multicolumn{9}{l}{Two-sided A-D statistics test values validated through 1000 bootstrap resampling.}\\
    \hline
    C/O$_{\text{D}}$        & 1.64  & 0.14 & 0.70 & 0.18 & 3.00  &    0.09       & 2.45  & 0.10\\
    C/O$_{\text{G}}$        &  --   &  --  &  --  &  --  & 5.72  &    0.02       & 5.31  & 0.03\\
    N/O$_{\text{G}}$        &  --   &  --  &  --  &  --  & 0.76  &    0.17       & 2.35  & 0.09\\
    A(C+N+O)$_{\text{G}}$   &  --   &  --  &  --  &  --  & 2.35  &    0.10       & 5.84  & 0.03\\
    Mg/Si                   & 0.87  & 0.17 & 1.95 & 0.13 & 31.17 & $\leq{0.001}$ & 26.15 & $\leq{0.001}$\\
    \hline
    \end{tabular}
    \tablefoot{The critical value of a 5$\%$ significance level for the A-D statistics is determined to be 1.961. The elemental C/O and N/O ratios in dwarfs are inconclusive, and no results are presented.}
   \end{table*}

The abundances of light elements in stars serve as critical constraints for studies on stellar yields, Galactic chemical evolution, and the chemical composition of exoplanets. Elemental ratios, such as the carbon-to-oxygen (C/O) and nitrogen-to-oxygen (N/O), are essential for understanding the structure, chemical composition, and potential migration history of exoplanets. Similarly, the magnesium-to-silicon (Mg/Si) ratio is particularly useful in determining their mineralogy.

We conducted two-sided K-S and A-D statistical tests to investigate potential differences in C/O, N/O, and Mg/Si elemental abundance ratios among stars with planets and the comparison sample. We segmented our analysis based on the masses of the planets orbiting these stars and categorised the host stars into two distinct groups: one group hosting low-mass planets ($M_{\text{p}}<30M_{\text{$\oplus$}}$) and the other hosting high-mass planets ($M_{\text{p}}>30M_{\text{$\oplus$}}$). For stars with multiple planets, the categorisation was as follows: if all planets in a system had masses exceeding 30$M_{\text{$\oplus$}}$, the star was classified in the group hosting high-mass planets, while stars with all planets having masses below 30$M_{\text{$\oplus$}}$ were classified in the group hosting low-mass planets. In cases where a star hosted planets of both mass ranges, it was included in both groups. Additionally, we focused on a majority of the stars in our sample that displayed thin disc-like kinematics. We then analysed whether the presence of planets had any impact on elemental composition within this subset.
   
We compiled the K-S statistics and $p$-values in Table~\ref{table:2} and A-D statistics, shown in Table~\ref{table:3}. In Fig.~\ref{fig:CDF}, we show the cumulative C/O, N/O and Mg/Si distributions. We also verified the distributions of the K-S and A-D statistics using 1,000 bootstrap resampling iterations. The $p$-value in Table~\ref{table:2} indicates the significance level of the K-S test, which represents the probability that the stars in our sample, whether they have planets or not, belong to the same population. The predetermined threshold is set at 5\%. A small $p$-value (i.e. $\le0.05$) in the K-S test indicates significant differences. Meanwhile, the A-D statistics in Table~\ref{table:3}, emphasize also potential distinctions in distribution tails. In the A-D test, the null hypothesis that stars with planets and comparison sample come from the same distribution is rejected if AD~$\geq$~AD$_{crit}$, where AD$_{crit}$ is the critical value and is equal to~1.961. 

\textbf{C/O ratio in low-mass planet host.} First, we take a look at the elemental C/O ratio in the low-mass planet host sample. In panel \textit{a} of Fig.~\ref{fig:CDF}, we observe that the cumulative C/O$_{\text{D}}$ ratio distributions for low-mass planet hosts (orange solid line) and the comparison sample (blue solid line) seem to exhibit different behaviours. The K-S test yielded a statistic value of 0.28 for the C/O in dwarfs (ref. as C/O$_{\text{D}}$), with corresponding $p$-value of 0.39 (there are 39\% chances the two samples come from the same distribution). These results suggest that there is no statistically significant difference in the distributions of elemental C/O$_{\text{D}}$ ratios (see Table~\ref{table:2}). This is further supported by bootstrap resampling ($p{\rm-value}=0.20$) and A-D statistics (0.42 and 1.64~$\leq$~AD$_{crit}$ in Table~\ref{table:3}). 

Following the initial analysis, we further investigated a subset of thin disc stars. When examining C/O$_{\text{D}}$ in low-mass planet hosts compared to the comparison sample, a K-S statistic of 0.21 and a high $p$-value of 0.81, suggests that there is no significant difference between the two distributions within the thin disc, the bootstrap resampling also confirms this insignificance. The A-D test results for the thin disc sample also show no significant difference, thus we cannot reject the null hypothesis because there is not sufficient evidence to say that C/O$_{\text{D}}$ in low mass thin disc planet-hosts is different from comparison stars.

\textbf{C/O and N/O ratio in high-mass planet host.}
The two statistical tests together with bootstrap resampling iterations show the same results for elemental C/O$_{\text{D}}$ ratios in high-mass planet hosts, suggesting that the samples could come from the same distribution. The elemental C/O$_{\text{G}}$ ratio in giant stars hosting high-mass planets show a significant difference with comparison sample (see panel \textit{b)} in Fig.~\ref{fig:CDF}) which is also confirmed by statistical results. For N/O$_{\text{~G}}$ ratio in high-mass hosts, K-S and A-D statistics show insignificant results while there are indications of moderate significance for the N/O$_{\text{~G}}$ ratio in the thin disc sample. Since the nitrogen (as well as C) is affected by mixing effects in giant stars, we proceed to analyse the total A(C+N+O) distribution in sec {\ref{subsec:CNO_in Giats_withPlanets}}.

\textbf{Mg/Si ratio in low- and high-mass planet host.} 
In panel~\textit{d}~of Fig.~\ref{fig:CDF}, we observe that stars with high-mass planets ($M_{\text{p}}>30M_{\text{$\oplus$}}$; red solid line) have, on average, significantly lower Mg/Si ratios than comparison stars (solid blue line) or stars with low-mass planets (solid orange line). These results are confirmed by two statistical tests. However, the low-mass planet sample needs more data.

\subsection{CNO in giant stars with planets} \label{subsec:CNO_in Giats_withPlanets}

   \begin{figure*}
    \centering
    \includegraphics[width=\hsize]{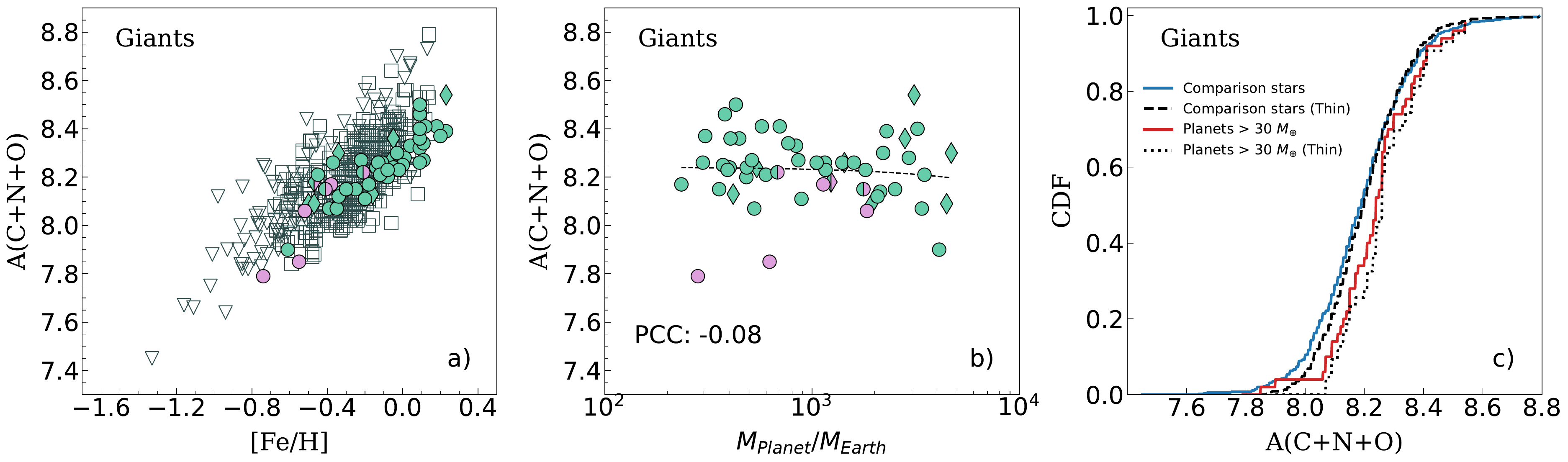}
    \caption{In panel \textit{a:} Distribution of A(C+N+O) abundances with respect to [Fe/H]. Panel \textit{b:} Same abundances versus planet mass for giant stars. Panel \textit{c:} Cumulative A(C+N+O) distributions for giant stars. All symbols have the same meaning as in Fig~\ref{fig:CDF}.} 
    \label{fig:A(C+N+O)}
   \end{figure*}
   
In Fig.~\ref{fig:CNOvsFeH(Age)}, we show the [C/Fe], [N/Fe], [O/Fe] abundances and C/O, N/O ratios as functions of [Fe/H] for giants stars only.  The planet-host giants are colour-coded by stellar ages (left panel) and stellar masses (right panel) determined in our study. We see that giant stars have different ages and masses. The carbon, nitrogen, and oxygen abundances in giant stars are affected by stellar evolution; however, the summed abundance of C+N+O conserves the initial conditions of star formation. Therefore, to potentially mitigate the influence of evolutionary effects, we also investigated the giant star sample hosting planets by the sum of C+N+O abundances. 

In panel \textit{a} of Fig.~\ref{fig:A(C+N+O)}, we show the distribution of A(C+N+O) abundances in giant stars versus metallicity [Fe/H], colour-coded as in other figures. We see that the total A(C+N+O) abundance increases with increasing [Fe/H], with a small scatter. In panel \textit{b} of Fig.~\ref{fig:A(C+N+O)}, we observe a total A(C+N+O) abundance versus planet mass and note an insignificant correlation with an increasing planetary mass (PCC$=-0.08$). In panel \textit{c} of Fig.~\ref{fig:A(C+N+O)}, we show the cumulative A(C+N+O) distribution for giant stars hosting giant planets (red solid line) in comparison to field stars (blue solid line). We see the two CDFs are separated, but the K-S and A-D statistics suggest no significant difference ($p-$values 0.24 and 0.11, respectively). Considering only the thin disc sample (represented by dotted and dashed lines), the separation remains evident. Two statistical tests support the possibility that the samples originate from different distributions, with a K-S p-value of 0.02 and an A-D p-value of 0.01. This conclusion is further reinforced by 1000 bootstrap resampling iterations, which yield a K-S p-value of 0.05 and an A-D p-value of 0.03. 

We also checked the A(C+N+O) abundances in dwarf stars versus metallicity [Fe/H] and planet mass, but due to the limited C, N, and O abundances determined for dwarf stars, we were unable to reach statistically significant conclusions. Further data collection and analysis will be necessary to extend these findings.

\subsection{Correlation between element abundances [X/Fe] and planet mass} \label{subsec:Abundance_vs_Mp} 

   \begin{figure*}
    \centering
    \includegraphics[width=0.9\hsize]{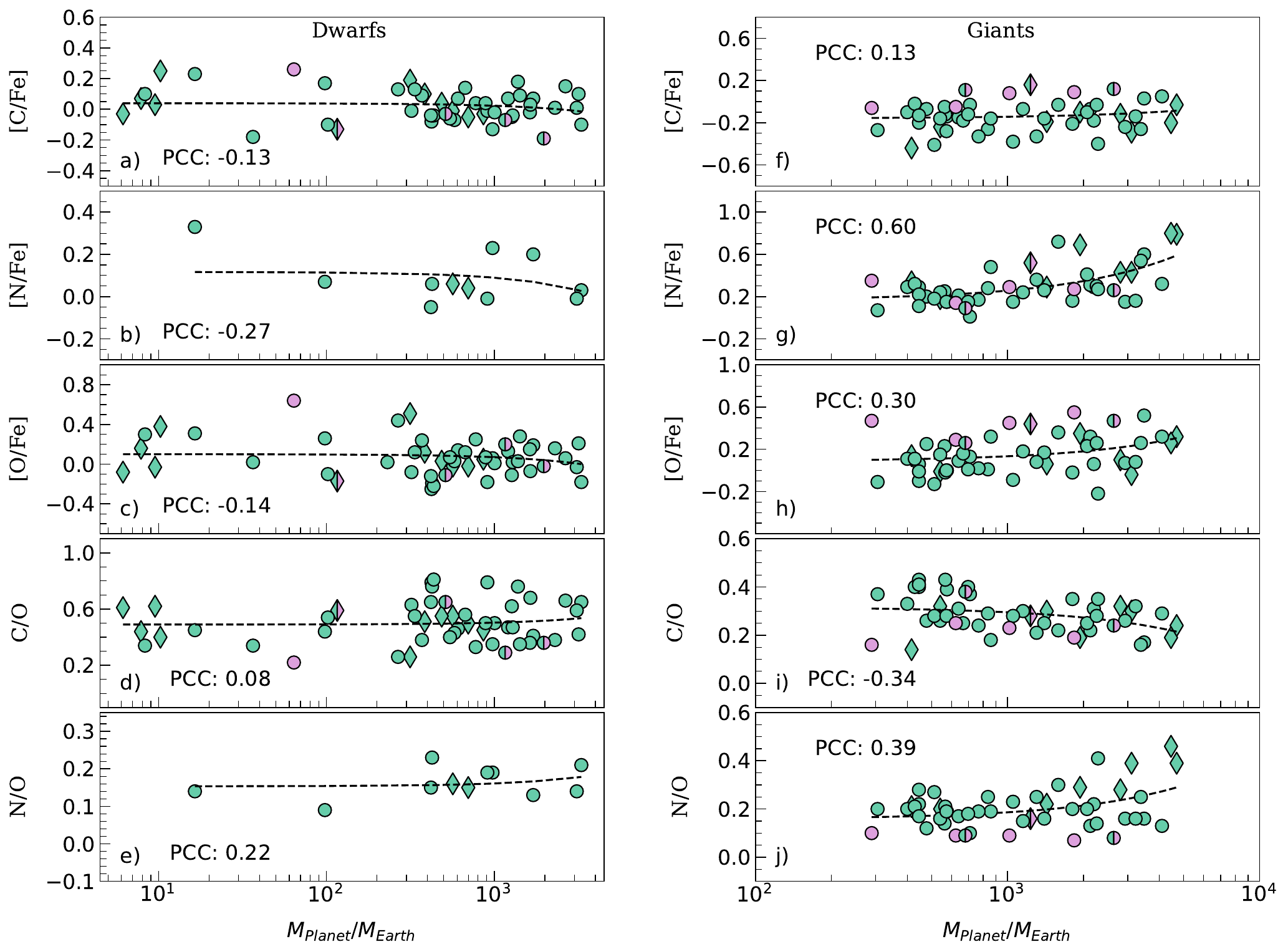}
    \caption{[C/Fe], [N/Fe], [O/Fe] abundances and C/O, N/O ratios as functions of planet masses for investigated dwarfs (left panel) and giants (right panel). Green circles and diamonds represent thin disc stars, while pink circles represent thick disc stars.}
    \label{fig:Pmass_CNO}
   \end{figure*}

   \begin{figure}
    \centering
    \includegraphics[width=0.9\hsize]{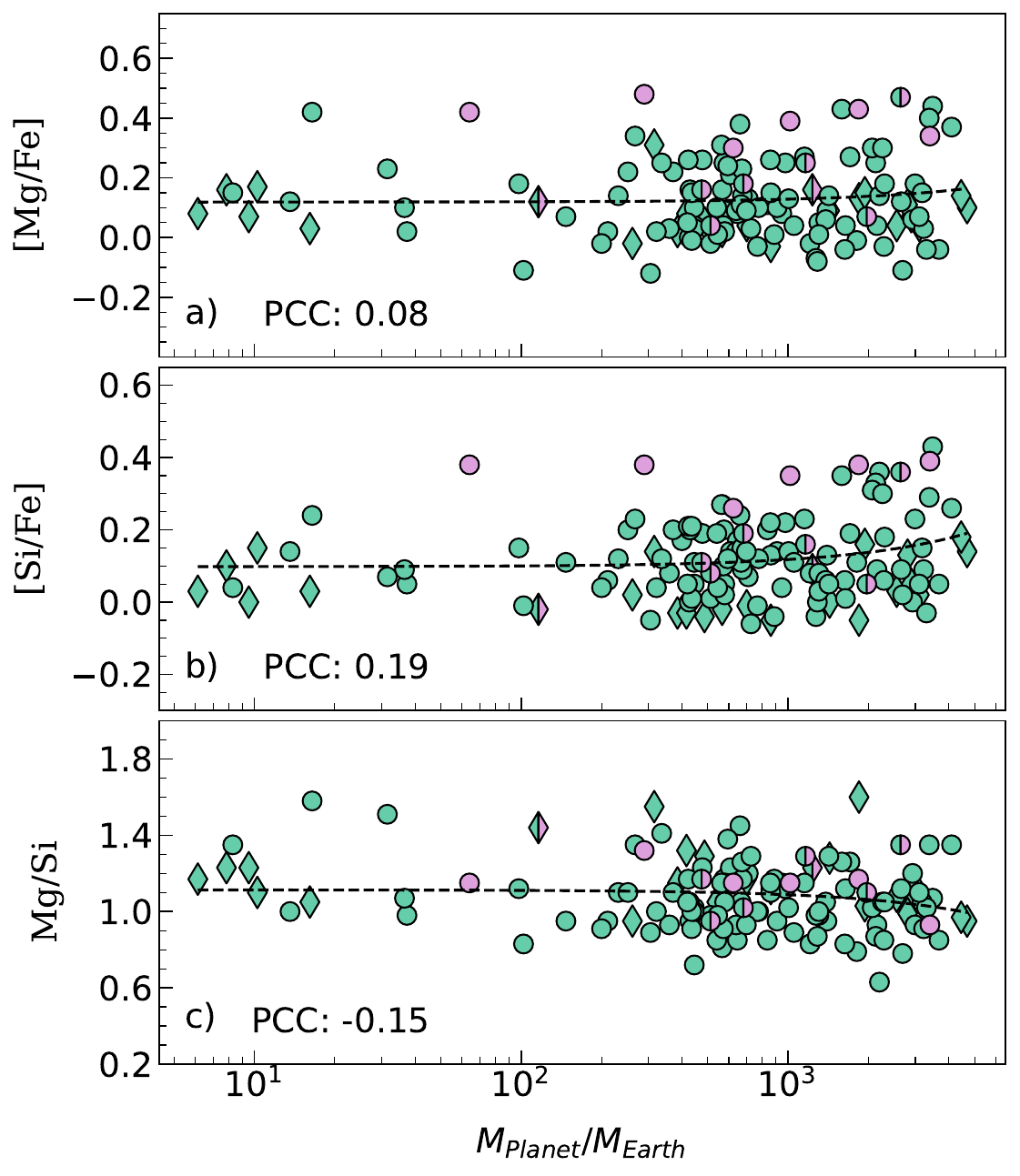}
    \caption{[Mg/Fe], [Si/Fe] abundances and their ratio Mg/Si as functions of planet masses for all investigated stars. All symbols have the same meaning as in Fig.~\ref{fig:Pmass_CNO}.}
    \label{fig:Pmass_MgSi}
   \end{figure}
   
Planets are believed to form through planetesimal accretion, pebble accretion, or a combination of both within protoplanetary discs surrounding young host stars. Therefore, we can expect that the present-day composition of a host star, if not significantly altered by stellar evolutionary effects, would primarily determine the compositions of its planets.

We investigated the correlations between the stellar abundances in our sample and the masses of their planetary companions. Following our previous methodology, we divided our host stars into dwarfs and giants. For our analysis, we excluded planets with masses greater than $\sim$13~M$_{J}$ where to potentially mitigate the effects of stellar evolution. This threshold is based on the limiting mass for the thermonuclear fusion of deuterium. The planetary masses, along with other parameters such as orbital periods and semi-major axes, were sourced from the literature, resulting in a non-homogeneous dataset. These parameters are summarised in Table~\ref{table:exoplanets}. For detailed information on the uncertainties associated with the planetary parameter values, please refer to the NASA Exoplanet Archive\footnote{For detailed information on the errors in the planetary parameter values, please refer to the NASA Exoplanet Archive: \url{https://exoplanetarchive.ipac.caltech.edu/}.}. 

In the case of multi-planetary systems (33 multi-planetary systems are included in our sample), we focused on the planet with the highest mass. Nevertheless, we also analysed the elemental abundance versus planet mass trends using the masses of all planets within these systems. Our analysis revealed no significant differences when including all planetary masses.

To explore the potential correlations, we conducted a linear regression analysis for the abundances as well as their ratios in relation to the masses of the planets. We computed linear fits using our dataset along with planet hosts from \cite{Tautvaisiene22}. We calculated the Pearson correlation coefficient (PCC) values to evaluate both the strength and direction of the linear relationship between the two variables.

Figure \ref{fig:Pmass_CNO} shows the distribution of [C/Fe], [N/Fe] and [O/Fe] element abundances as well as the elemental C/O and N/O ratios as functions of planet masses separately for investigated dwarf and giant stars. We observe a weak negative correlation among the three elemental abundances for dwarf stars, which is consistent with our findings analysing smaller sample size \cite{Tautvaisiene22}. \cite{Suarez-Andres17} showed similar results for dwarfs underscoring a consistent absence of correlation in the case of carbon abundance. In giants, we found a weak carbon abundance correlation with planet mass (PCC=0.13). For nitrogen, we observed a strong positive relationship in giant stars that host high-mass planets. For oxygen, we see a moderate positive relationship in the direction of high mass planet-hosts.

In the case of the C/O ratio, we found a weak positive correlation between C/O and planet mass in dwarfs. On the contrary, for giant hosts, there is a moderate negative C/O slope with a PCC of $-0.34$ for stars hosting high-mass planets. 

The linear fit of the N/O ratio versus planet mass for dwarf hosts indicates a weak positive correlation (PCC value of 0.22). However, more data on the low-mass planet hosts are needed to confirm this trend. Among giant stars, the calculated PCC value for the N/O ratio is 0.39 indicating a moderate positive correlation between the N/O ratio and planetary mass. 

Figure \ref{fig:Pmass_MgSi} shows the distribution of [Mg/Fe] and [Si/Fe] element abundances and Mg/Si ratio as functions of planet masses for all investigated stars. The symbols have the same meaning as in Fig.~\ref{fig:Pmass_CNO}. Similar to C, N, and O abundance results, we found a linear trend between magnesium and silicon abundances and planet mass. There is a negative Mg/Si slope (PCC$=-0.15$) towards the stars hosting high-mass planets.

\subsection{Stellar age as a function of planet mass} \label{subsec:Age_vs_Mp}

   \begin{figure}  
    \centering
    \includegraphics[width=\hsize]{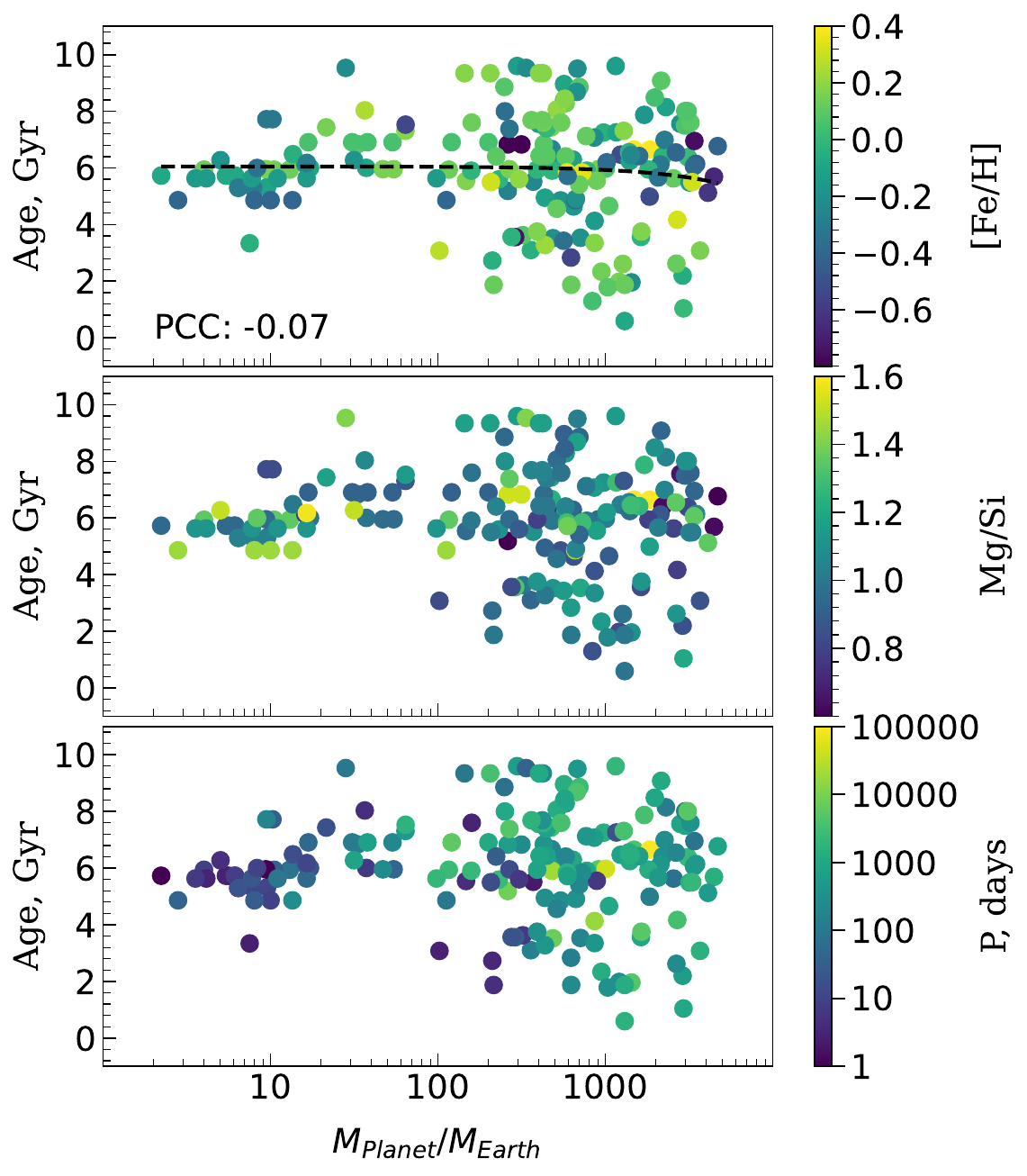}
    \caption{Distribution of stellar age versus planet mass colour-coded by [Fe/H] (top panel), Mg/Si (middle panel) \& orbital period of the planets (bottom panel). Refer to the text for more information.}
    \label{fig:AgevsPmass}
   \end{figure}

Recently, \citet{Swastik24} analysed the ages for a large number of exoplanet-hosting stars and found that the median ages of the stars hosting small planets are higher compared to stars hosting giant planets (M$_{P}$ $\ge$ 0.3M$_J$). Authors suggested that the later chemical enrichment of the galaxy by the iron-peak elements, largely produced from Type Ia supernovae, may have paved the way for the formation of gas giants. Furthermore, within the giant planet population, stars hosting hot Jupiters (orbital period $\le  10$~days) are found to be younger compared to the cool and warm Jupiters planet hosts (orbital period $>$10 days), implying that hot Jupiters could be the youngest systems to emerge in the progression of planet formation. This motivated us to look for possible correlations between planet parameters (mass, period) and stellar age in our sample. 

In Fig.~\ref{fig:AgevsPmass}, we show planet mass versus stellar age distributions for dwarfs and giant stars, colour-coded by [Fe/H], Mg/Si and planet orbital period (in days). We find a flat trend with a hint of decreasing stellar age with planet mass (PCC$=-0.07$), but more data is needed to confirm this trend. Furthermore, we see that stars hosting low-mass short-period planets are on average older (and more metal-poor). Long-period giant planets are found around old and young stars. In addition, we see that the short-period and younger giant planet hosts have on average lower elemental Mg/Si ratios (see also Fig.~\ref{fig:Pmass_MgSi}).

\section{Summary and conclusions}\label{sec:conclusions}

Using high-resolution spectroscopy, we investigated a combined sample of 149 stars with planets, comprising 124 newly observed stars and 25 stars from our previous works. We focused on the abundance analysis of light elements (carbon, nitrogen, oxygen, magnesium, silicon), that are important constraints for stellar yields, Galactic chemical evolution, and exoplanet chemical composition studies.

Stellar atmospheric parameters were uniformly derived using the classical equivalent width approach. Based on the ${\rm log}~g$ values, our sample of planet-hosts has been divided into 83 main sequence stars, referred to in this paper as dwarfs, and 66 evolved giant stars. ${\rm Log}~g$ values for giants in our sample varied between 1.3~to~3.5~dex with a mean of $2.8\pm0.6$~dex, and for dwarfs between $\geq$3.6~to~4.6~dex with a mean of $4.2\pm0.2$~dex. $T_{\rm eff}$ varied from 3998 to 6675~K and [Fe/H] from $-$0.76 to 0.35~dex, with a mean of $-$$0.08\pm0.20$~dex for all investigated stars.

We also determined Galactic space velocities, \textit{U}, \textit{V}, \textit{W}, and orbital parameters: mean galactocentric distance, $R\mathrm{_{mean}}$, maximum vertical distance from the Galactic plane, $|z_{\rm max}|$, and eccentricity, $e$ for newly observed sample of stars. $R\mathrm{_{mean}}$ values for this sample varied between 5.79 to 9.85 kpc, with a mean of 7.80~kpc. The maximum $|z_{\rm max}|$ value reached 1.90~kpc with the average of 0.30~kpc. We also estimated the ages and masses of the stars as well as disc probability ratios to group our stars into thin or thick disc stars. Stars which fall in between thin and thick discs are referred to in this paper as in-between stars. The stellar ages varied between 0.59 to 9.60~Gyr and we have a distinctive group of 109 thin and 7 thick-disc stars along with 8 in-between stars. The separation of the remaining 25 host stars into thin and thick discs was taken from \citep{Tautvaisiene22}.

Element abundances are derived using the spectral synthesis method by comparing observed spectra with modelled spectra. No systematic differences in derived abundances between dwarfs and giants were found when plotted against atmospheric parameters. Furthermore, a separate analysis of the abundance patterns of carbon (C), nitrogen (N), and oxygen (O) relative to iron, as well as the C/O, N/O and Mg/Si abundance ratios, and the combined absolute C+N+O abundance for dwarfs and giants, was conducted across the full range of metallicity. 

The results indicate lower C and O abundances in dwarf planet-hosts at super-solar metallicities than in comparison sample. However, at lower metallicities, there is no significant difference in C and O abundances between planet hosts and comparison samples. The C/O distributions are, on average, higher in metal-rich dwarfs compared to metal-poor dwarfs with no significant differences between planet hosts and comparison samples, as indicated by the K-S and A-D tests. The nitrogen abundance in metal-rich dwarf stars tends to follow solar values, with an increase in nitrogen abundance observed at sub-solar metallicities. The N/O ratio seems to increase with [Fe/H] without significant differences between planet hosts and comparison samples, as indicated by the K-S and A-D tests. We acknowledge the necessity for additional homogeneous nitrogen abundance data to analyse nitrogen trends in planet-hosting stars.

C, N and O abundances in giant planet hosts are consistent with the comparison sample. However, giant planet hosts exhibit less scatter in carbon and oxygen abundances and tend to have lower carbon levels at a given metallicity. Additionally, the C/O ratio in planet-hosting giants is lower, and the N/O ratio increases with [Fe/H] without significant differences between planet hosts and comparison samples, as indicated by the K-S and A-D tests.

To potentially reduce the impact of stellar evolutionary effects, we have also analysed the combined absolute abundances of C, N and O, referred to as A(C+N+O) in giant stars hosting planets. Our analysis indicates that stars with higher metal content generally exhibit higher C+N+O abundances. Comparing the A(C+N+O) abundances in giant planet hosts with the comparison sample, we found that the CDFs for A(C+N+O)  are closely aligned, indicating no clear separation between them. Moreover, K-S and A-D test results further confirm that this difference is not statistically significant. Within thin disc stars, a potential separation in the chemical properties of stars hosting giant planets compared to comparison stars is indicated. However, further studies with larger samples are necessary to confirm these trends.
 
Analysing the abundance patterns of magnesium (Mg) and silicon (Si) relative to iron, as well as the Mg/Si ratio between stars hosting planets and comparison sample, reveals a notable overabundance of Mg and Si in planet-hosting stars, particularly on the lower-metallicity side. These results suggest the critical role of these elements in the process of planet formation. Our results on Mg and Si also align with the Galactic chemical evolution as we observed higher abundances of Mg and Si in metal-poor stars compared to metal-rich stars. Furthermore, the observed Mg/Si ratio appears to be, on average, slightly lower in stars with planets compared to field stars at a given [Fe/H]. Additionally, as indicated by the K-S and A-D tests, the elemental Mg/Si ratio in stars hosting high-mass planets is significantly different than the comparison sample.

We also explored the correlations between the abundances of stars and the masses of their planetary companions, leading to the following conclusions:

\begin{itemize}
 \item In dwarf stars, a weak negative correlation was observed between the abundance ratios [C/Fe], [N/Fe], and [O/Fe] and planet mass. In contrast, for giant stars, a weak positive correlation was observed between [C/Fe] and planet mass (PCC = 0.13) while a strong positive correlation was observed between [N/Fe] and planet mass (PCC = 0.60). The [O/Fe] ratio exhibits a moderate positive relation with planet mass (PCC = 0.30).
 
 \item A comparative analysis showed that giants have slightly lower [C/Fe] but higher [N/Fe] and [O/Fe] abundances than dwarfs, due to evolutionary mixing processes. Among dwarf stars, the C/O ratio shows a weak positive correlation with planet mass (PCC = 0.08). Conversely, in giant stars hosting high-mass planets, the C/O ratio demonstrates a moderate negative correlation (PCC = $-0.34$). The N/O ratio shows a weak positive correlation with planet mass in both dwarfs (PCC = 0.22) and a moderate positive correlation in giants (PCC = 0.39).

 \item The observed abundances of [Mg/Fe] and [Si/Fe] demonstrate weak positive correlations with planet masses, evidenced by PCC of 0.08 and 0.19, respectively. This trend is more pronounced in stars hosting high-mass planets. Conversely, the Mg/Si ratio shows a weak negative correlation with planet mass, with a PCC of $-0.15$, a trend that is similarly more evident in stars with high-mass planets.
\end{itemize}

We looked for possible correlations between stellar parameters, specifically the metallicity ([Fe/H]), and magnesium-to-silicon ratio (Mg/Si), as well as planet parameters such as the mass and orbital period, with respect to stellar ages within our dataset. 
We found a flat trend between planet mass and stellar age, showing a hint of stellar age with planet mass correlation (PCC$=-0.07$), but more data are needed to confirm this trend.

Additionally, older stars of our sample with lower metallicity ([Fe/H]) more often host low-mass, short-period planets, aligning with theoretical models indicating earlier planet formation in older stars. Differences in Mg/Si ratios also correlate with the stellar age and planet orbital period, with younger stars hosting short-period giant planets and showing lower Mg/Si ratios.


\vspace{2mm}
\hspace{5mm}
\textit{Facility}: NASA Exoplanet Archive.

\vspace{2mm}
\textit{Softwares}: Astropy \citep{AstropyCollaboration}, DAOSPEC \citep{Stetson08}, MOOG \citep{Sneden73}, galpy \citep{Bovy15}, TURBOSPECTRUM \citep{Alvarez98}, UniDAM \citep{Mints17}.
   
\begin{acknowledgements}
We thank the referee, Tamara Mishenina, for insightful comments and valuable suggestions. E.S., A.D., R.M., {\v S}.M. and G.T. acknowledge funding from the Research Council of Lithuania (LMTLT, grant No. P-MIP-23-24). This research has made use of the NASA Exoplanet Archive, which is operated by the California Institute of Technology, under contract with the National Aeronautics and Space Administration under the Exoplanet Exploration Program. We also acknowledge the use of the SIMBAD database, operated at CDS, Strasbourg, France. Additionally, we appreciate Vilnius University's Moletai Astronomical Observatory for granting us observation time for this project. The observing time was partially funded by the Europlanet Telescope Network programme of the Europlanet 2024 Research Infrastructure project. Europlanet 2024 RI has received funding from the European Union's Horizon 2020 research and innovation programme under grant agreement No 871149.
\end{acknowledgements}

\bibliographystyle{aa}
\bibliography{sharma.bbl}

\appendix 
\section{Figures}

\begin{figure*}
    \centering
    \includegraphics[width=\hsize]{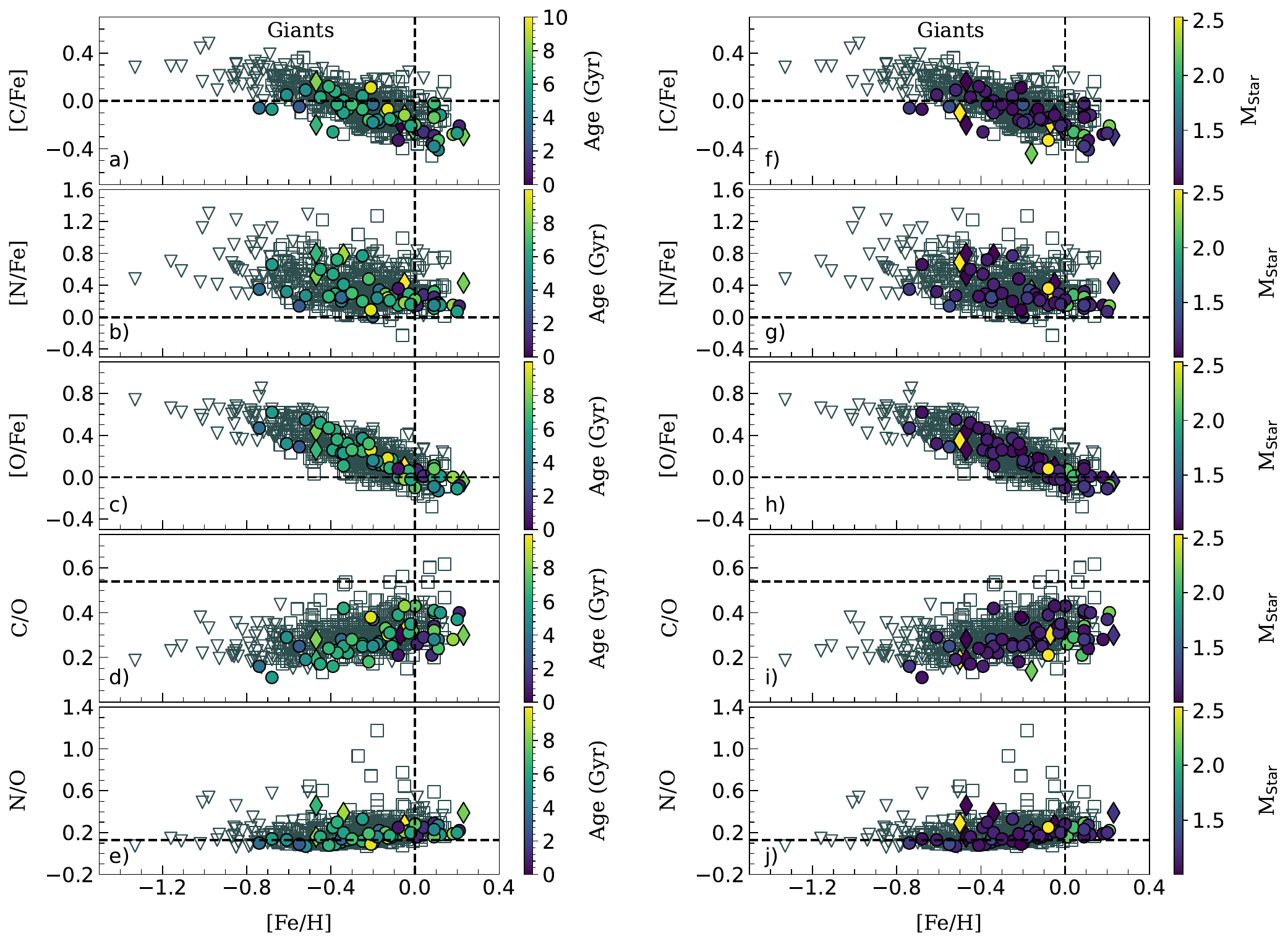}
    \caption{[C/Fe], [N/Fe], and [O/Fe] abundances and C/O and N/O ratios, plotted as functions of [Fe/H] for observed giant stars. Similar to Fig~\ref{fig:CNOvsFeH}, the results from comparison stars are taken from \cite{Tautvaisiene22} indicated in grey empty squares (thin disc) and triangles (thick-disc). The investigated stars are colour-coded by stellar ages (left panel) and masses (right panel).}
    \label{fig:CNOvsFeH(Age)}
   \end{figure*}

In Fig.~\ref{fig:CNOvsFeH(Age)}, we show the [C/Fe], [N/Fe], and [O/Fe] abundances and C/O and N/O ratios plotted as functions of [Fe/H] for giants stars only. The planet-host giant stars are now colour-coded by stellar ages (left panel) and stellar masses (right panel). We see that on average, the metal-rich stars are the youngest and most massive. 

\begin{figure*}
    \centering
    \includegraphics[width=\hsize]{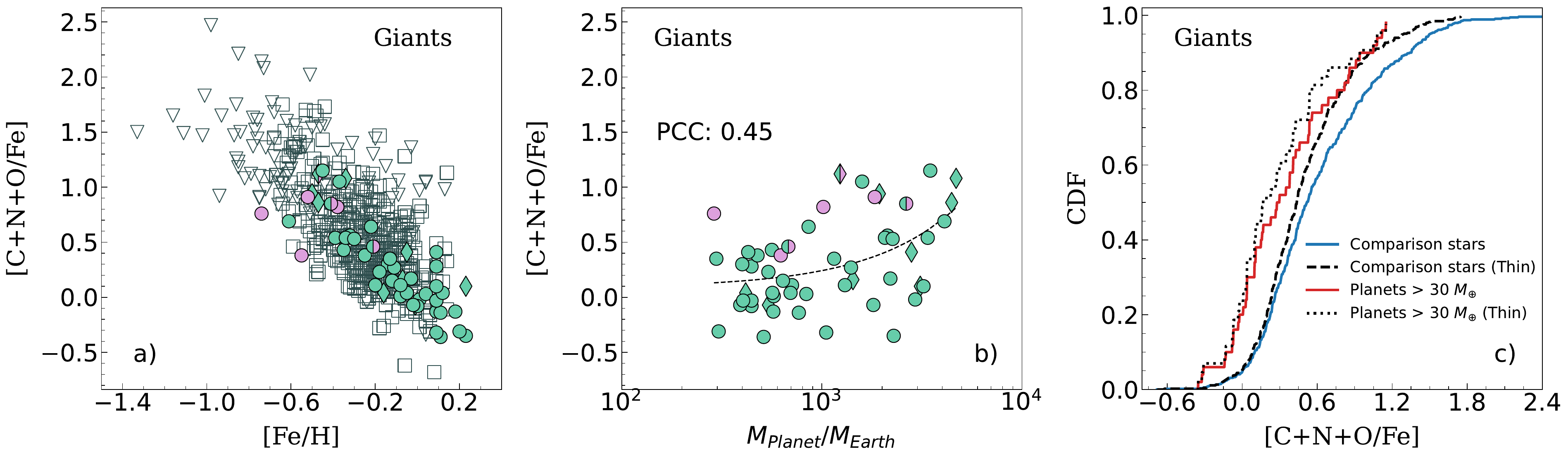}
    \caption{In panel \textit{a:} Distribution of [C+N+O/Fe] abundances with respect to [Fe/H]. Panel \textit{b:} Same abundances versus planet mass for giant stars. Panel \textit{c:} Cumulative [C+N+O/Fe] distributions for giant stars. All symbols have the same meaning as in Fig~\ref{fig:CDF}.}
    \label{fig:[C+N+O/Fe]}
   \end{figure*}

In panel \textit{a}~of~Fig.~\ref{fig:[C+N+O/Fe]}, we show the distribution of [C+N+O/Fe] abundances with respect to [Fe/H]. Panel \textit{b} displays the same abundances versus planet mass for giant stars, while panel \textit{c} shows the cumulative [C+N+O/Fe] distributions for giant stars. All symbols have the same meaning as in Fig~\ref{fig:CDF}. The results from panel \textit{a} demonstrate that the combined [C+N+O/Fe] ratio decreases with increasing metallicity [Fe/H], as anticipated by Galactic chemical evolution. The metal-poor (thick disc) stars have on average higher combined [C+N+O/Fe] abundance. The results from panel \textit{b} shows that combined [C+N+O/Fe] abundance in thin- and thick disc stars increases with planet mass.

\section{Data tables}

Table~\ref{table:Results} provides a comprehensive summary of our analysis results. This includes the main atmospheric parameters, ages, kinematic and orbital properties, and chemical composition of the stars along with associated uncertainties. For those interested in exploring the complete dataset, the entire table is available in a machine-readable form at CDS.

In Table~\ref{table:exoplanets}, we presented key planetary parameters and elemental ratios for the stellar systems analysed in this study. The table includes parameters such as planet mass, orbital period and orbital semi-major axis along with the C/O, N/O and Mg/Si abundance ratios of their host stars. For comprehensive access to the data, including the complete list of parameters and the full set of planetary systems considered in this study, the entire table is available through the Centre de Données astronomiques de Strasbourg (CDS).

\onecolumn
 \begin{longtable}{llll}
 \caption{Contents of the machine-readable table available online at CDS.}
 \label{table:Results}\\
 \hline
 \hline
 Col & Label & Units & Explanations\\
 \hline
 1	& Host TYC ID        & ---          & Tycho-2 catalogue identification\\
 2	& Teff               & K    	    & Effective temperature\\
 3	& e\_Teff            & K    	    & Error in effective temperature\\
 4	& Logg               & [cm/s$^{2}$] & Stellar surface gravity\\
 5	& e\_Logg            & [cm/s$^{2}$] & Error in stellar surface gravity\\
 6	& [Fe/H]             & dex   	    & Metallicity\\
 7	& e\_[Fe/H]          & dex   	    & Error in metallicity\\
 8	& Vt                 & km/s         & Microturbulence velocity\\
 9	& e\_Vt              & km/s         & Uncertainty in microturbulence velocity\\
 10	& Vrad               & km/s         & Radial velocity\\
 11	& e\_Vrad            & km/s         & Uncertainty in radial velocity\\
 12	& Age                & Gyr          & Stellar age\\
 13	& e\_Age             & Gyr          & Uncertainty in stellar age\\
 14	& U                  & km/s         & Heliocentric space velocity $U$\\
 15	& e\_U               & km/s         & Uncertainty in heliocentric space velocity $U$\\
 16	& V                  & km/s         & Heliocentric space velocity $V$\\
 17	& e\_V               & km/s         & Uncertainty in Heliocentric space velocity $V$\\
 18	& W                  & km/s         & Heliocentric space velocity $W$\\
 19	& e\_W               & km/s         & Uncertainty in Heliocentric space velocity $W$\\
 20	& d                  & kpc 	        & Stellar distance\\
 21	& R$_{\rm mean}$     & kpc    	    & Mean Galactocentric distance\\
 22	& e\_R$_{\rm mean}$  & kpc 	        & Uncertainty in mean Galactrocentric distance\\
 23	& z$_{\rm max}$      & kpc    	    & Maximum distance from Galactic plane\\
 24	& e\_z$_{\rm max}$   & kpc    	    & Uncertainty in maximum distance from Galactic plane\\
 25	& {\it{e}}           & ---    	    & Orbital eccentricity\\
 26	& e\_{\it{e}}        & ---  	    & Uncertainty in orbital eccentricity\\
 27	& [C/H]    	         & dex    	    & Carbon abundance\\
 28	& e\_[C/H]  	     & dex     	    & Uncertainty in carbon abundance\\
 29	& [N/H]    	         & dex    	    & Nitrogen abundance\\
 30	& e\_[N/H]  	     & dex     	    & Uncertainty in nitrogen abundance\\
 31	& [O/H]    	         & dex    	    & Oxygen abundance\\
 32	& e\_[O/H]  	     & dex     	    & Uncertainty in oxygen abundance\\
 33	& [Mg/H]    	     & dex    	    & Magnesium abundance\\
 34	& e\_[Mg/H]  	     & dex     	    & Uncertainty in magnesium abundance\\
 35	& [Si/H]    	     & dex    	    & Silicon abundance\\
 36	& e\_[Si/H]  	     & dex     	    & Uncertainty in silicon abundance\\
 37	& C/O    	         & dex    	    & Carbon to Oxygen abundance ratio\\
 38  & N/O                & dex          & Nitrogen to Oxygen abundance ratio\\
 39	& Mg/Si    	         & dex    	    & Magnesium to Silicon abundance ratio\\
 40	& TD/D               & --- 	        & Thick disk-to-thin disk probability ratio\\
 41	& Thin $\vert$ Thick & ---     	    & Chemical attribution to the Galactic subcomponent\\
 \noalign{\smallskip}
 \hline
 \end{longtable}
 \centering

   \begin{table*}[h]
    \caption{Table displaying planetary parameters and derived elemental ratios for stars in our sample. The parameters are taken from NASA exoplanet archive. The elemental abundance ratios are determined in this work. Full table is available in machine-readable form at CDS.}
    \label{table:exoplanets}
    \centering
    \renewcommand{\tabcolsep}{1.85mm}
    \renewcommand{\arraystretch}{1.35}
    \begin{tabular}{l c c c c c c c c} 
    \hline
    \hline
    {\raisebox{-1.5ex}[0cm][0cm]{Host TYC ID}}  &  { \raisebox{-1.5ex}[0cm][0cm]{Planet}}  &  Planet Mass, $M_\mathrm{p}$  &  Orbital Period, \textit{P}  &  Semi Major Axis, \textit{a}  &  {\raisebox{-1.5ex}[0cm][0cm]{C/O}}  &  {\raisebox{-1.5ex}[0cm][0cm]{N/O}}  &  {\raisebox{-1.5ex}[0cm][0cm]{Mg/Si}}  &  Ref.$^{(1)}$   \\
    &&  ($M_\mathrm{Earth}$)  &  (days)  &  (au)  &  \\
    \hline
     
    1949-2012-1	 &  55 Cnc b  &	267.0  &  14.65171    &  0.11620  &  0.79  &  0.19  &  0.95  &  [1]\\
                 &  55 Cnc c  &	54.4   &  44.38270    &  0.24320  &&&&  [1]\\
                 &  55 Cnc d  &	909.0  &  4820.00000  &  5.54000  &&&&  [1]\\
                 &  55 Cnc e  &	9.4    &  0.73654     &  0.01583  &&&&  [1]\\
                 &  55 Cnc f  &	46.9   &  260.98000   &  0.79200  &&&&  [1]\\
    \noalign{\smallskip}
    
    ...  & ...  &  ...  &  ...  &  ...  &  ...  &  ...  &  ...  &  ...  \\
    \hline
    \end{tabular}
    \flushleft{\bf{References.}} $^{(1)}$ Planetary parameters references: [1] - \citet{Rosenthal21}. 
   \end{table*}


\end{document}